\documentclass[11pt]{article}

\usepackage[reqno]{amsmath}
\usepackage{epsfig}
\usepackage{array}
\usepackage{float}

\usepackage{a4}
\usepackage{epsfig}
\usepackage{a4wide}
\def\be{\begin{equation}}
\def\ee{\end{equation}}
\def\gs{\mathrel{
   \rlap{\raise 0.511ex \hbox{$>$}}{\lower 0.511ex \hbox{$\sim$}}}}
\def\ls{\mathrel{
   \rlap{\raise 0.511ex \hbox{$<$}}{\lower 0.511ex \hbox{$\sim$}}}}

\newcommand{\ba}{\begin{array}{c}}
\newcommand{\baz}{\begin{array}{cc}}
\newcommand{\bad}{\begin{array}{ccc}}
\newcommand{\bea}{\begin{equation} \begin{array}{c}}
\newcommand{\eea}{ \end{array} \end{equation}}
\newcommand{\ea}{\end{array}}
\newcommand{\D}{\displaystyle}

\def\gtap{\mathrel{
   \rlap{\raise 0.511ex \hbox{$>$}}{\lower 0.511ex \hbox{$\sim$}}}}
\def\ltap{\mathrel{
   \rlap{\raise 0.511ex \hbox{$<$}}{\lower 0.511ex \hbox{$\sim$}}}}
\newcommand{\deltaatm}{\mbox{$\Delta  m^2_{\mathrm{atm}}$}}
\newcommand{\deltasol}{\mbox{$ \Delta  m^2_{\odot}$}}
\newcommand{\utre}{\mbox{$|U_{\mathrm{e} 3}|$}}
\newcommand{\betabeta}{\mbox{$(\beta \beta)_{0 \nu}  $}}

\newcommand{\meff}{\mbox{$\left|  < \! m  \! > \right| \ $}}
\newcommand{\hbeta}{$\mbox{}^3 {\rm H}$ $\beta$-decay }

\newcommand{\deltatre}{\mbox{$ \ \Delta m^2_{32} \ $}}
\newcommand{\deltadue}{\mbox{$ \ \Delta m^2_{21} \ $}}

\newcommand{\uetre}{\mbox{$ \ |U_{\mathrm{e} 3}|^2  \ $}}
\newcommand{\me}{\mbox{$ m_{\bar{\nu}_{e}}$}}

\newcommand{\deltaatmmax}{\mbox{$(\Delta  m^2_{\mathrm{atm}})_{ \! \mbox{}_{\mathrm{MAX}}} \ $}}
\newcommand{\deltaatmmin}{\mbox{$(\Delta  m^2_{\mathrm{atm}})_{ \! \mbox{}_{\mathrm{MIN}}} \ $}}

\newcommand{\utremax}{\mbox{$|U_{\mathrm{e} 3}|^2_{ \! \mbox{}_{\mathrm{MAX}}}$ }}
\newcommand{\utremin}{\mbox{$|U_{\mathrm{e} 3}|^2_{ \! \mbox{}_{\mathrm{MIN}}} $ }}

\newcommand{\uuno}{\mbox{$|U_{\mathrm{e} 1}|^2$}}

\begin{document}

\hfill{Ref. SISSA 60/2002/EP}
\rightline{DO-TH 02/12}
\rightline{August 2002}
\rightline{hep-ph/0208XXX}

\begin{center}
{\bf On the CP Violation Associated with Majorana Neutrinos \\
and Neutrinoless Double-Beta Decay }

\vspace{0.4cm}

S. Pascoli~$^{a,b)}$, \hskip 0.2cm
S. T. Petcov~$^{a,b)}$
\footnote{Also at: Institute of Nuclear Research and
Nuclear Energy, Bulgarian Academy of Sciences, 1784 Sofia, Bulgaria}
~and~
W. Rodejohann~$^{c)}$ 

\vspace{0.2cm}

{\em $^{a)}$Scuola Internazionale Superiore di Studi Avanzati, 
I-34014 Trieste, Italy\\
}
\vspace{0.2cm}   
{\em $^{b)}$Istituto Nazionale di Fisica Nucleare, 
Sezione di Trieste, I-34014 Trieste, Italy\\
}
\vspace{0.2cm}   
{\em $^{c)}$Department of Physics, University of Dortmund, 
Germany\\
}
\end{center}

\begin{abstract}
   Assuming 3-$\nu$ mixing and massive Majorana neutrinos,
we analyze the possibility of establishing 
the existence of CP-violation  
associated with Majorana neutrinos
in the lepton sector by
i) measuring of the effective Majorana mass
\meff in neutrinoless double beta decay with 
a sufficient precision and ii) by measuring of, or
obtaining a stringent upper limit on, the 
lightest neutrino mass $m_1$. Information on $m_1$
can be obtained in the \hbeta experiment KATRIN and
from astrophysical and cosmological observations.
Proving that the indicated 
CP-violation takes place requires, in particular, 
a relative experimental error on the measured value of 
\meff not bigger than 20\%,
a ``theoretical uncertainty'' in the value of
\meff due to an imprecise knowledge of the 
corresponding nuclear matrix elements
smaller than a factor of 2, a value of 
$\tan^2\theta_{\odot} \gtap 0.55$,
and values of the relevant Majorana
CP-violating phases typically 
within the intervals of $\sim (\pi/2 - 3\pi/4)$ and
$\sim (5\pi/4 - 3\pi/2)$. 
  
\end{abstract}

\newpage
\section{Introduction}
\vspace{-0.2cm}

\hskip 1.0truecm The recent results 
of the SNO solar neutrino experiment \cite{SNO2,SNO3}
(see also \cite{SNO1}) 
provided further strong evidences 
for oscillations or transitions
of the solar $\nu_e$ into active neutrinos 
$\nu_{\mu (\tau)}$ (and/or antineutrinos $\bar{\nu}_{\mu (\tau)}$).
These evidences become even stronger 
when the SNO data are combined with the
data obtained in the other solar neutrino experiments,
Homestake, Kamiokande, SAGE, GALLEX/GNO and 
Super-Kamiokande \cite{SKsol,Cl98}.
As the two-neutrino oscillation analyzes
of the solar neutrino data show (see, e.g., \cite{SNO2}),  
the latter favor  the large mixing angle (LMA) 
MSW $\nu_e \rightarrow \nu_{\mu (\tau)}$ transition
solution with $\Delta m^2_{\odot} \sim 5\times 10^{-5}~{\rm eV^2}$
and $\tan^2\theta_{\odot} \sim 0.33$, $\tan^2\theta_{\odot} < 1$,
where $\Delta m^2_{\odot}$ and $\theta_{\odot}$ are
the neutrino mass squared difference and mixing
angle which control the 
solar neutrino transitions.
Strong evidences for oscillations of 
atmospheric neutrinos have been obtained in the 
Super-Kamiokande experiment \cite{SKatm00}.
The atmospheric neutrino data, 
as is well known,
is best described in terms of
dominant $\nu_{\mu} \rightarrow \nu_{\tau}$
($\bar{\nu}_{\mu} \rightarrow \bar{\nu}_{\tau}$)
oscillations with 
$|\Delta m^2_{\mathrm{atm}}| \sim (2.5 - 3.0)\times 10^{-3}~{\rm eV^2}$.
 
    The explanation of the solar and 
atmospheric neutrino data in terms of 
neutrino oscillations requires
the existence of 3-neutrino mixing
in the weak charged lepton current 
(see, e.g., \cite{BGG99,P99}):
%%%%%%%%%%%%%%%%%%%
\begin{equation}
\nu_{l \mathrm{L}}  = \sum_{j=1}^{3} U_{l j} \, \nu_{j \mathrm{L}}~.
\label{3numix}
\end{equation}
%%%%%%%%%%%%%%%%%%%
\noindent Here $\nu_{lL}$, $l  = e,\mu,\tau$,
are the three left-handed flavor 
neutrino fields,
$\nu_{j \mathrm{L}}$ is the 
left-handed field of the 
neutrino $\nu_j$ having a mass $m_j$
and $U$ is the Pontecorvo-Maki-Nakagawa-Sakata (PMNS)
neutrino mixing matrix \cite{BPont57,MNS62}. 
If the neutrinos with definite mass $\nu_j$
are Majorana particles, 
the process of neutrinoless double-beta  
(\betabeta-) decay, 
$(A,Z) \rightarrow (A,Z+2) + e^{-} + e^{-}$,
$(A,Z)$ and $(A,Z+2)$ being 
initial and final state nuclei,
will be allowed (for reviews see, 
e.g., \cite{BiPet87,ElliotVogel02}).
For Majorana neutrinos $\nu_j$
with masses not exceeding few MeV,
the dependence of the 
\betabeta-decay amplitude 
on the neutrino mass and mixing parameters
is confined to one factor --- 
the effective Majorana mass \meff$\!\!$, 
which can be written in the form 
(see, e.g., \cite{BiPet87}):
%%%%%%%%%%%%%%%%%%%%
\begin{equation}
\meff \!\! = \left| m_1 |U_{\mathrm{e} 1}|^2 
+ m_2 |U_{\mathrm{e} 2}|^2~e^{i\alpha_{21}}
 + m_3 |U_{\mathrm{e} 3}|^2~e^{i\alpha_{31}} \right|
\label{effmass2}
\end{equation}
%%%%%%%%%%%%%%%%%%%%
\noindent where 
$\alpha_{21}$ and $\alpha_{31}$ 
are the two Majorana CP-violating phases
\footnote{We assume that $m_j > 0$ and that
the fields of the 
Majorana neutrinos $\nu_j$ 
satisfy the Majorana condition:
$C(\bar{\nu}_{j})^{T} = \nu_{j},~j=1,2,3$,
where $C$ is the charge conjugation matrix.}
\cite{BHP80,Doi81}.
If CP-invariance holds, 
one has \cite{LW81,BNP84}
$\alpha_{21} = k\pi$, $\alpha_{31} = 
k'\pi$, where $k,k'=0,1,2,...$. In this case 
%%%%%%%%%%%%%%%%%%%%
\begin{equation}
\eta_{21} \equiv e^{i\alpha_{21}} = \pm 1,~~~
\eta_{31} \equiv e^{i\alpha_{31}} = \pm 1 ,
\label{eta2131}
\end{equation}
%%%%%%%%%%%%%%%%%%%%
\noindent represent the relative 
CP-parities of the neutrinos 
$\nu_1$ and $\nu_2$, and 
$\nu_1$ and $\nu_3$, respectively. 

 One can express 
\cite{SPAS94,BGKP96,BGGKP99,BPP1,Paramm1}
the masses $m_2$ and $m_3$ entering into 
eq.\ (\ref{effmass2}) for \meff in terms of 
\deltasol{} and \deltaatm, measured in the solar and 
atmospheric neutrino experiments, and $m_1$, 
while $|U_{\mathrm{e} j}|^2$, 
$j=1,2,3$, are related
to the mixing angle
which controls the solar 
$\nu_e$ transitions $\theta_{\odot}$,
and to the lepton mixing parameter $\sin^2\theta$
limited by the data from the CHOOZ
and Palo Verde experiments \cite{CHOOZ,PaloV}.
Within the convention $m_1 < m_2 < m_3$
we are going to use 
in what follows, one has $\deltaatm \equiv \Delta m^2_{31}$,
where $\Delta m^2_{jk} \equiv m_j^2 - m_k^2$,
and $m_3 = \sqrt{m_1^2 + \deltaatm}$.
For \deltasol{}  there are two possibilities, 
$\deltasol \equiv \Delta m^2_{21}$ 
and $\deltasol \equiv \Delta m^2_{32}$,
corresponding respectively to two different types
of neutrino mass spectrum ---  
with normal and with inverted hierarchy.
In the first case one has
$m_2 = \sqrt{m_1^2 + \deltasol}$,
$|U_{\mathrm{e} 1}|^2 = \cos^2\theta_{\odot} (1 - |U_{\mathrm{e} 3}|^2)$, 
$|U_{\mathrm{e} 2}|^2 = \sin^2\theta_{\odot} (1 - |U_{\mathrm{e} 3}|^2)$,
and  $|U_{\mathrm{e} 3}|^2 \equiv \sin^2\theta$,
while in the second 
$m_2 = \sqrt{m_1^2 + \deltaatm - \deltasol}$,
$|U_{\mathrm{e} 2}|^2 = \cos^2\theta_{\odot} (1 - |U_{\mathrm{e} 1}|^2)$, 
$|U_{\mathrm{e} 3}|^2 = \sin^2\theta_{\odot} (1 - |U_{\mathrm{e} 1}|^2)$,
and  $|U_{\mathrm{e} 1}|^2 \equiv \sin^2\theta$.
Thus, given 
\deltasol, \deltaatm, $\theta_{\odot}$ and
$\sin^2\theta$, \meff depends, in general,
on the lightest neutrino mass $m_1$,
on the two Majorana CP-violating phases 
$\alpha_{21}$ and $\alpha_{31}$ and on the 
``discrete ambiguity'' related to the two possible
types of neutrino mass spectrum.
In the case of quasi-degenerate (QD) 
neutrino mass spectrum,
$m_1 \cong m_2 \cong m_3$, $m_1^2 \gg \deltaatm,\deltasol$,
~\meff essentially does not depend on
\deltaatm{} and \deltasol, and 
the two possibilities, 
$\deltasol \equiv \deltadue$
and $\deltasol \equiv \deltatre \!\!$, 
lead to the same predictions for  
\footnote{This statement is valid 
as long as there are no independent
constraints on the  CP-violating phases
$\alpha_{21}$ and $\alpha_{31}$ 
which enter into the expression for \meff$\!\!$.}
\meff$\!\!$.
 
 The observation of \betabeta-decay
will have fundamental implications 
for our understanding of the
elementary particle interactions. It would imply, 
in particular, that the electron lepton charge 
$L_e$ and the total lepton charge $L$ are not conserved 
and can change by two units in the latter,
and would suggest that the massive neutrinos are
Majorana particles.
Under the general and plausible assumptions of 
3-$\nu$ mixing and massive Majorana neutrinos, 
\betabeta-decay generated 
only by the (V-A) charged current weak interaction 
via the exchange of the three Majorana neutrinos,
and neutrino oscillation explanation of the solar 
and atmospheric neutrino data, 
which will be assumed to hold throughout this study,
the observation of \betabeta-decay
\footnote{Evidences for \betabeta-decay
taking place with a rate corresponding to
 $0.11 \ {\rm eV} \leq  \meff \!\! \leq  0.56$ eV
(95\% C.L.) are claimed to 
have been obtained in \cite{Klap01}. The
results announced in \cite{Klap01} have been 
criticized in \cite{bb0nu02}.}  
can give unique information on 
the type of the neutrino mass spectrum 
and on the lightest neutrino mass, i.e., 
on the absolute scale of neutrino masses
\cite{BGGKP99,BPP1,Paramm1,WR00,BPP2,PPW,WR0302,PPSNO2bb,bb0nuMassSpec1,Weiler2001}.
One of the important implications of the latest
results of the solar neutrino experiments,
notably of SNO, which show that $\tan^2\theta_{\odot} < 1$ 
and, e.g., $\cos2\theta_{\odot} \gtap 0.26$ at 99.73\% C.L. \cite{SNO2}, 
regard the predictions for \meff$\!\!$. 
The fact that $\cos2\theta_{\odot}$ 
is significantly different from zero leads to 
\cite{PPSNO2bb} (see also \cite{BPP1,PPW,WR0302})
the existence of significant lower 
bounds on \meff (exceeding 0.01 eV)
in the cases of neutrino mass spectrum 
with inverted hierarchy (IH) and
of the quasi-degenerate (QD) type, and 
of a stringent upper bound 
(smaller than 0.01 eV) 
if the neutrino mass spectrum is 
with normal hierarchy (NH).
Using, e.g., the best fit values of
\deltasol, \deltaatm, $\cos2\theta_{\odot}$
and $\sin^2\theta$, obtained in the analyzes
of the solar and atmospheric neutrino, and 
CHOOZ data in \cite{SNO2,Gonza3nu},
one finds respectively 
for the three types of spectra \cite{PPSNO2bb}:
$\meff \!\! \gtap 2.8\times 10^{-2}~{\rm  eV}$ (IH),
$\meff \!\! \gtap 0.06~{\rm  eV}$ (QD) and
$\meff \!\! \ltap 2.0\times 10^{-3}~{\rm eV}$ (NH).
At 90\% C.L. the indicated
lower and upper bounds read, respectively:
$\meff \!\! \gtap 1.5\times 10^{-2}~{\rm  eV}$ (IH),
$\meff \!\! \gtap 0.25~{\rm  eV}$ (QD) and
$\meff \!\! \ltap 6.0\times 10^{-3}~{\rm eV}$ (NH).
The quoted lower bounds 
are in the range of the sensitivity of 
currently operating 
and planned 
\betabeta-decay experiments (see further).
These results imply, in particular, that  
a measured value of $\meff \!\! \neq 0$
(or an experimental
upper limit on \meff$\!\!$) of the order of
${\rm few} \times 10^{-2}$ eV 
can provide unique constraints on,
or even can allow one to determine, 
the type of the neutrino mass spectrum in the case 
the massive neutrinos are Majorana particles; 
it can provide also a significant upper limit 
on the mass of the lightest
neutrino $m_1$ \cite{PPW,WR0302,PPSNO2bb}.  
Information on absolute values of neutrino masses
in the range of interest
can also be obtained in the 
\hbeta neutrino mass experiment KATRIN \cite{KATRIN}
and from cosmological and astrophysical 
data (see, e.g., refs.~\cite{Weiler2001,Hu99}).

    Rather stringent upper
bounds on \meff have been obtained in the 
$^{76}$Ge experiments 
by the Heidelberg-Moscow collaboration \cite{76Ge00}, 
$ \meff \!\! < 0.35~{\rm eV}$ ($90\%$ C.L.), 
and by the IGEX collaboration \cite{IGEX00},
$\meff \!\! < (0.33 \div 1.35)~{\rm eV}$ ($90\%$ C.L.).
Taking into account a factor of 3 uncertainty
in the calculated value of the corresponding 
nuclear matrix element, we get for the upper limit
found in \cite{76Ge00}:  $\meff \!\! < 1.05$ eV.
Considerably higher sensitivity to the value of 
$\meff$ is planned to be 
reached in several $\betabeta$-decay experiments
of a new generation. 
The NEMO3 experiment \cite{NEMO3}, 
which began to take data in July of 2002, 
and the cryogenics detector CUORICINO 
\cite{CUORE} to be operative in the second half of 2002,
are expected to reach a sensitivity to values of 
$\meff \!\! \sim 0.2~$eV.
Up to an order of magnitude better sensitivity,
i.e., to $\meff \!\! \cong 2.7\times 10^{-2}$ eV,
$1.5\times 10^{-2}~$eV, $5.0\times 10^{-2}~$eV,
$2.5\times 10^{-2}$ eV and $3.6\times 10^{-2}$ eV
is planned to be achieved 
in the CUORE \cite{CUORE}, GENIUS \cite{GENIUS},
EXO 
\cite{EXO}, 
MAJORANA \cite{Maj} and MOON \cite{MOON}
experiments
\footnote{The quoted sensitivities 
correspond to values of the relevant nuclear matrix
elements taken from ref.\ \cite{SMutoKK90}.}, 
respectively.

   In what regards the \hbeta experiments, 
the currently existing most stringent upper 
bounds on the electron (anti-)neutrino mass  
$m_{\bar{\nu}_e}$ were obtained in the
Troitzk~\cite{MoscowH3} and Mainz~\cite{Mainz} 
experiments and read
$m_{\bar{\nu}_e} < 2.2$ eV\@.
The KATRIN \hbeta experiment \cite{KATRIN}
is planned to reach a sensitivity  
to  $m_{\bar{\nu}_e} \sim 0.35$ eV.

  In the present article we discuss 
the possibility of establishing the existence of 
CP-violation in the lepton sector due to the
Majorana CP-violating phases by measuring \meff$\!\!$.
The fundamental problem of CP-violation 
in the lepton sector is one of the 
most challenging future frontiers in the 
studies of neutrino mixing.
It was noticed in \cite{BGKP96} (see also \cite{BGGKP99})
that in the case of a {\it large mixing 
angle solution of the solar neutrino problem},
the observation of \betabeta-decay 
combined with data on the neutrino masses from
\hbeta experiments could provide, in principle,
unique information on the CP-violation
due to the Majorana CP-violating phases.
As a more detailed study 
showed \cite{BPP1}, information 
on the CP-violation of interest,
and if CP-invariance holds --- on the
relative CP-parities of the massive 
Majorana neutrinos, 
could be obtained as well
from a measurement of \meff
supplemented by information on the type of the
neutrino mass spectrum (or the lightest    
neutrino mass $m_1$). 
The problem of interest was studied 
in detail for the QD neutrino mass spectrum
(taking into account the 
relevant nuclear matrix element
uncertainties) in \cite{WR00}.
Further analysis 
\footnote{Aspects of the phenomenology 
of the effects of CP-violation in
\betabeta-decay were discussed
also, e.g., in \cite{bb0nuCP1}.}
showed \cite{PPW}
that  a measurement of \meff alone
could exclude the possibility of 
both Majorana CP-violating phases 
$\alpha_{21}$ and $\alpha_{31}$
being equal to zero. However,
such a measurement cannot rule out
without additional input
that the two phases take different 
CP-conserving values.
The additional input needed for
establishing CP-violation
could be, e.g., the measurement of 
neutrino mass $m_{\bar{\nu}_e}$ in 
\hbeta experiment
KATRIN \cite{KATRIN}, or the 
cosmological determination of
the sum of the three neutrino masses \cite{Hu99},
$\Sigma = m_1 + m_2 + m_3$,
or a derivation of a sufficiently 
stringent upper limit on $m_1$ (or $\Sigma$).
It was also pointed out in \cite{PPW}
that the possibility of finding CP-violation
``requires quite accurate measurements''
of \meff and, say, of $m_{\bar{\nu}_e}$, 
``and holds only for a limited range of 
values of the relevant parameters''.
The aim of the present paper is
to quantify these requirements,
and to better determine the ranges 
of values of the parameters 
in question, which could allow to 
detect CP-violation due to the
Majorana CP-violating phases.
We discuss also the 
requirement the possibility
of establishing CP-violation imposes 
on the uncertainty in the values
of the \betabeta-decay nuclear matrix elements.    
Let us add  that at present 
no viable alternative to the measurement of 
\meff for getting information 
about the Majorana CP-violating phases 
$\alpha_{21}$ and $\alpha_{31}$
exists (see, e.g., \cite{WRJPhys}), or can be foreseen 
to exist in the next $\sim 8$  years.

   The problem of detecting CP-violation 
associated with Majorana neutrinos by measuring $\meff$ 
and $m_{\bar{\nu}_e}$ (or $\Sigma$)
was discussed recently also in ref.\ \cite{BargerSNO2bb}.
The authors of \cite{BargerSNO2bb}, after making a certain
number of assumptions about the 
experimental and theoretical developments
in  the field of interest that might occur by 2020
\footnote{It is supposed in \cite{BargerSNO2bb}, in particular, 
that \meff will be measured with a 25\% (1 s.d.) error and that
the uncertainty in the \betabeta-decay nuclear matrix elements 
will be reduced to a factor of 2.}, 
claim to have shown ``once and for all
that it is impossible to detect CP-violation from 
\betabeta-decay in the foreseeable future.''
We have strong doubts that it is 
possible to foresee with certainty
all the scientific and technological developments
relevant to the problem of interest,
which will take place in the next $\sim (10 - 18)$ years.
Correspondingly, the approach 
we follow in the present work is ``orthogonal'' 
to that adopted in \cite{BargerSNO2bb}:
here we make an attempt to determine the 
conditions under which 
CP-violation might be detected   
from a measurement of 
\meff and  $m_{\bar{\nu}_e}$ (or $\Sigma$),
or of \meff and a sufficiently 
stringent upper limit on $m_1$ or $\Sigma$.
%%%%%%%%%%%%%%%%%%%%%%%%%%%%%%%%%%%%%%%%%%%%%%%%%%%%%%%%%%%%%%%
\vspace{-0.4cm}
\section{The Neutrino Mass and Oscillation Data and the 
Predictions for the Effective Majorana Mass \meff}
\vspace{-0.2cm}
%%%%%%%%%%%%%%%%%%%%%%%%%%%%%%%%%%%%%%%%%%%%%%%%%%%

\hskip 1.0cm  As we have seen, the predicted value 
of \meff depends in the 
case of 3-neutrino mixing of interest on:
i) the value of the lightest neutrino mass $m_1$,
ii) $\Delta m^2_{\odot}$ and $\theta_{\odot}$, 
iii) 
$\deltaatm$, and 
iv) the  lepton mixing angle $\theta$ which is
limited by the CHOOZ and Palo Verde 
experiments \cite{CHOOZ,PaloV}.
Given the indicated parameters,
the value of \meff depends strongly 
on the type of the
neutrino mass spectrum, as well as 
on the values of the two
Majorana CP-violating phases,
$\alpha_{21}$ and $\alpha_{31}$ 
(see eq.\ (\ref{effmass2})),
present in the lepton mixing matrix.

    The possibility of detecting
of CP-violation due to the
Majorana CP-violating phases 
$\alpha_{21}$ and $\alpha_{31}$
if \meff is found to be nonzero in the 
\betabeta-decay experiments of the next 
generation, depends crucially  
on the precision with which 
$m_1$ (or $\Sigma$), \deltasol, $\theta_{\odot}$,
\deltaatm, $\sin^2\theta$ and \meff 
will be measured. It depends also crucially, 
as we shall see, on the values of $m_1$ or 
$\Sigma$, of
$\theta_{\odot}$ and of \meff$\!\!$.
Actually, the accuracy of measurement of 
\meff in the next generation of \betabeta-
decay experiments, given their sensitivity
limits of $\sim (1.5 - 5.0)\times 10^{-2}~{\rm eV}$,
depends on the value of \meff$\!\!$.   
If only an upper limit on $m_1$ 
will be obtained, 
the possibility we are discussing
will depend on how stringent 
this upper limit is.
In what regards the dependence 
of \meff on the type of the
neutrino mass spectrum (normal versus 
inverted hierarchy),
if the latter will not be determined in
neutrino oscillation experiments
(see, e.g., \cite{SPMPiai01,AMMS00}),
the measurement of $\meff \!\! \neq 0$ itself  
in the next generation of \betabeta-decay
experiments could provide this information
through the value of \meff found \cite{PPSNO2bb}.

   The value of $m_1$ can be measured 
by the KATRIN experiment \cite{KATRIN} 
if the neutrino mass spectrum is of 
the QD type. In this case
$m_1 \cong m_2  \cong m_3 \cong m_{\bar{\nu}_e}$.
Given the currently allowed regions of values of 
\deltasol{}  and \deltaatm{} (see further), we have QD spectrum
for $m_{1,2,3} \cong m_{\bar{\nu}_e} > 0.20$ eV.
The KATRIN detector is designed to have a
1 s.d.\ error of 0.08 eV$^2$ on a measured value of
$m_{\bar{\nu}_e}^2$. The most stringent 
upper limit on $m_{\bar{\nu}_e} \cong m_1$, 
which can be reached
in this experiment, is 0.35 eV. The KATRIN experiment 
is expected to start in 2007.

   The sum of neutrino masses $\Sigma$ can be 
determined by using data on the weak lensing
of galaxies by large scale structure,
and data on the cosmic microwave background (CMB)
from the MAP and PLANCK experiments, 
with an estimated (1 s.d.) error of 0.04 eV \cite{Hu99}.
The latter represents the estimated best precision that 
can possibly be achieved in the cosmological
determination of $\Sigma$.
If only an upper limit 
of $\sim (0.10 - 0.15)$ eV on $\Sigma$ 
will be obtained, that would strongly disfavor 
(if not rule out) the QD spectrum, while a measured 
value of $\meff \!\! \gtap 0.03$ eV would rule out
the NH spectrum \cite{PPSNO2bb}.
Given \deltasol{}  and \deltaatm,
the indicated upper limit on $\Sigma$ 
could be used to derive a rather stringent upper limit 
on $m_1$: using $\Sigma < 0.15~(0.12)$ eV and
the current best fit values of
$\deltasol \cong 5.0\times 10^{-5}~{\rm eV^2}$
and  $\deltaatm \cong 3.0\times 10^{-3}~{\rm eV^2}$,
we get in the case of IH spectrum:
$m_1 < 0.03~(0.01)$ eV. An upper limit
on $m_1$ of the order of (0.010 -- 0.025) eV 
might be of crucial importance
for establishing CP-violation due to the 
Majorana CP-violating phases \cite{BPP1,PPW,WR0302}.

    The analysis of the solar neutrino data 
\cite{SNO2,SNO3,SNO1,SKsol,Cl98},
including the latest SNO results,
in terms of the hypothesis of 
$\nu_e \rightarrow \nu_{\mu (\tau)}$
oscillations/transitions of the solar $\nu_e$ 
shows \cite{SNO2} (see also, e.g., 
\cite{BargerSNO2,FogliSNO2}) 
that the data favor the 
LMA MSW solution with $\deltasol > 0$
and $\tan^2\theta_{\odot} < 1$.
The LOW solution of the 
solar neutrino problem
with transitions into 
active neutrinos is only allowed at
approximately 99.73\% C.L. \cite{SNO2};
there do not exist other solutions
at the indicated confidence level.
In the case of the LMA solution, 
the range of values of 
$\Delta m^2_{\odot}$
found in \cite{SNO2} at 99.73\% C.L. reads:
%%%%%%%%%%%%%%%%%%%%%%%
\begin{equation}
{\rm LMA~MSW}:~~~~~~2.2\times 10^{-5}~{\rm eV^2} 
\ltap \Delta m^2_{\odot} 
\ltap 2.0\times 10^{-4}~{\rm eV^2}~~~~~(99.73\%~{\rm C.L.}).
\label{dmsolLMA}
\end{equation}
%%%%%%%%%%%%%%%%%%%%%%%
% 
\noindent The best fit value of $\Delta m^2_{\odot}$
obtained in \cite{SNO2} is 
$(\Delta m^2_{\odot})_{\mathrm{BF}} = 5.0\times 10^{-5}~{\rm eV^2}$.
The mixing angle $\theta_{\odot}$
was found in the case of the LMA solution
to lie in an interval which 
at 99.73\% C.L. is given by \cite{SNO2}
%%%%%%%%%%%%%%%%%%%%%%%
\begin{equation}
{\rm LMA~~MSW}:~~~~~~~~~~~~~~~~~
0.26 \ltap \cos2\theta_{\odot} \ltap 0.64~~~~~~~(99.73\%~{\rm C.L.}).~~~~~~~~~
\label{thLMA}
\end{equation}
%%%%%%%%%%%%%%%%%%%%%%%%%%%
%
\noindent The best fit value of
$\cos2\theta_{\odot}$ 
in the LMA solution region is given by    
$(\cos2\theta_{\odot})_{\mathrm{BF}} = 0.50$.

   Similar results have been obtained, e.g., 
in \cite{BargerSNO2,FogliSNO2}; in particular,
the minimal allowed values of
$\cos 2\theta_{\odot}$ in the LMA solution region
found in \cite{SNO2} and in \cite{BargerSNO2} 
at 99.73\% C.L. practically coincide.
The minimal allowed value of
$\cos 2\theta_{\odot}$ 
obtained in \cite{FogliSNO2}
at  99.73\% C.L. is 0.10.
The best fit values of 
$\cos 2\theta_{\odot}$ 
found in \cite{SNO2,BargerSNO2,FogliSNO2}
coincide, while that of \deltasol{}
obtained in \cite{FogliSNO2},
$(\deltasol)_{\mathrm{BF}} 
\cong 5.5\times 10^{-5}~{\rm eV^2}$
is only slightly larger than the value
found in \cite{SNO2,BargerSNO2}.

  In the  two-neutrino $\nu_{\mu} \rightarrow \nu_{\tau}$
($\bar{\nu}_{\mu} \rightarrow \bar{\nu}_{\tau}$)
oscillation analysis of the atmospheric neutrino
data performed in \cite{SKatm00} 
the following best fit value of \deltaatm{} 
was obtained: $(\deltaatm)_{\mathrm{BF}} 
\cong 2.5\times 10^{-3}~{\rm eV^2}$.
At 99.73\% C.L. \deltaatm{} was found to lie 
in the interval: $(1.5 - 5.0)\times 10^{-3}~{\rm eV^2}$. 

   A 3-$\nu$ oscillation analysis of the CHOOZ data 
showed \cite{BNPChooz}, e.g., 
that for $\deltasol \ltap 10^{-4}~{\rm eV^2}$,
the limits on $\sin^2\theta$ practically coincide with
those derived in the 2-$\nu$ oscillation analysis
in ref.\ \cite{CHOOZ}. Combined 3-$\nu$ oscillation analyzes of the 
solar neutrino, Super-Kamiokande atmospheric neutrino
and CHOOZ data were performed in \cite{Gonza3nu,Fogli3nuSNO2KamL}
under the assumption of $\deltasol \ll \deltaatm$
(see, e.g., \cite{BGG99,P99,ADE80}).
For the best fit values of
\deltaatm{} and $\sin^2\theta$ the authors 
of \cite{Gonza3nu} and \cite{Fogli3nuSNO2KamL} 
obtained, $(\deltaatm)_{\mathrm{BF}} 
\cong 3.1\times 10^{-3}~{\rm eV^2}$,
$(\sin^2\theta)_{\mathrm{BF}} \cong 0.005$,
and $(\deltaatm)_{\mathrm{BF}} 
\cong 2.7\times 10^{-3}~{\rm eV^2}$,
$(\sin^2\theta)_{\mathrm{BF}} \cong 0$, 
respectively. It was found in \cite{Fogli3nuSNO2KamL}, 
in particular, that $\sin^2\theta < 0.05$ at 99.73\% C.L. 

   If $\Delta m^2_{\odot} \cong (2.5 - 10.0)\times 10^{-5}~{\rm eV^2}$,
which is favored by the solar neutrino data,
the KamLAND experiment taking data at present
will be able to measure
$\Delta m^2_{\odot}$  with an 1 s.d.\ error of
$\sim (3 - 5)\%$ 
(see, e.g., refs.\ \cite{Carlos01} and 
\cite{Fogli3nuSNO2KamL,VitoAKamL02} and the 
articles quoted therein). 
Combining the data from the solar neutrino experiments 
and from KamLAND would permit to determine
$\tan^2\theta_{\odot}$ with a high precision as well: 
the estimated (1 s.d.) error on $\tan^2\theta_{\odot}$ is 
\footnote{We thank C. Pe\~na-Garay for 
clarifications on this point.}
$\sim 5\%$ \cite{Carlos01}. 
 
   Similarly, if $\deltaatm$ lies in the interval 
$\deltaatm \cong (2.0 - 5.0)\times 10^{-3}~{\rm eV^2}$, 
as is suggested by the current 
atmospheric neutrino data \cite{SKatm00}, 
its value will be determined with a 
$\sim 10\%$ error (1 s.d.)  by the MINOS experiment
\cite{MINOS} which is scheduled to start 
in December of 2004.   
Somewhat better limits on $\sin^2 \theta$ than 
the existing one can be obtained in the 
MINOS experiment \cite{MINOS} as well. 
Various options are being currently discussed
(experiments with off-axis neutrino beams, more precise
reactor antineutrino and long base-line experiments, etc.,
see, e.g., \cite{MSpironu02}) of how to improve
by at least an order of magnitude, i.e., 
to values of $\sim 0.005$ or smaller, 
the sensitivity to $\sin^2\theta$. 

   All the indicated developments
are expected to take place 
within the next $\sim (7 - 8)$ years, i.e., by 2010. 
We will assume in what follows
that the problem of measuring 
or tightly constraining $\sin^2\theta$ will also be  
resolved within the indicated period. 
We will also assume that by 2010 one or more
\betabeta-decay experiments of the next generation
will be operative, and that at least the physical range
of variation of the values of the relevant
\betabeta-decay nuclear matrix elements 
will be unambiguously determined.
Since all discussed future
experiments (except the 10 ton version of
GENIUS \cite{GENIUS}) have a sensitivity to 
$\meff \!\! > 0.01$ eV, while in the case of NH 
neutrino mass spectrum \meff is predicted to be smaller
than 0.01 eV, we will not consider
the possibility of finding CP-violation 
for the NH spectrum (for a discussion of this possibility
see, e.g., \cite{PPW,WR0302}).

  We begin with a  general analysis 
the aim of which is to determine what is the
maximal uncertainty in the value of \meff due to the
imprecise knowledge of the corresponding nuclear matrix
elements, which might still allow one to
find CP-violation associated with Majorana neutrinos.
This is followed 
by results on the problem of finding the CP-violation
of interest, derived by a simplified 
error analysis
of prospective input data on 
$m_{\bar{\nu}_e}$ or $\Sigma$,
$\tan^2\theta_{\odot}$, \meff$\!\!$, etc.
The effects of the nuclear matrix element
uncertainty on the results of the statistical 
analysis are also considered.
%%%%%%%%%%%%%%%%%%%%%%%%%%%%%%%%%%%%%%%%%%%%%%%%%%%%%%%%%%%%%%%%
\vspace{-0.4cm}
\section{% Searching for 
Finding CP-Violation Associated with Majorana Neutrinos}
\vspace{-0.2cm}
%%%%%%%%%%%%%%%%%%%%%%%%%%%%%%%%%%%%%%%%%%%%%%%%%%%
%%%%%%%%%%%%%%%%%%%%%%%%%%%%%%%%%%%%%%%%%%%%%%%%%%%%%%%%%%%%%%%%
\vspace{-0.1cm}
\subsection{General Constraints on the Nuclear Matrix Element Uncertainty}
\vspace{-0.1cm}
%%%%%%%%%%%%%%%%%%%%%%%%%%%%%%%%%%%%%%%%%%%%%%%%%%%
%%%%%%%%%%%%%%%%%%%%%%%%%%%%%%%%%%%%%%%%%%%%%
\vspace{-0.1cm}
\subsubsection{Inverted Neutrino Mass Hierarchy: 
$\deltasol \equiv \deltatre$, $m_1 < 0.02 \ {\rm eV}$}
\vspace{-0.2cm}
%%%%%%%%%%%%%%%%%%%%%%%%%%%%%%%%%%%%%%%%%%%%%

\hskip 1.0cm   If $\deltasol = \Delta m^2_{32}$, one has 
\cite{BGGKP99,BPP1}:
%%%%%%%%%%%%%%%%%%%%%%%%%%%%%%%%%%%%%%%%%%%%%%
\begin{eqnarray}
\meff  =  & \Big| m_1 |U_{\mathrm{e}1}|^2  
  + 
 \sqrt{m_1^2 + \deltaatm \! \!  - \deltasol}
 \cos^2 \theta_\odot (1 - |U_{\mathrm{e}1}|^2)  e^{i\alpha_{21}}
 \nonumber \\ 
& + 
 \sqrt{ m_1^2 + \deltaatm} \sin^2 \theta_\odot 
 (1 - |U_{\mathrm{e}1}|^2) e^{i\alpha_{31} }
 \Big|
\label{eqmassinvhierarchy01} \\ 
&  \hspace{-4.7cm}  \simeq   \Big|
 \sqrt{ \deltaatm }
 \big( \cos^2 \theta_\odot   
+  \sin^2 \theta_\odot 
  e^{i\alpha_{32} }\big)
 \Big| = \sqrt{ \deltaatm } \sqrt{ 1 - 
\sin^22\theta_{\odot}~\sin^2\frac{\alpha_{32}}{2}}~,
\label{eqmassinvhierarchy02} 
\end{eqnarray}
%%%%%%%%%%%%%%%%%%%%%%%%%%%%%%%%%%%%%%%%%%%%%%%
%
where $\alpha_{32} = \alpha_{31} - \alpha_{21}$.
In eq.~(\ref{eqmassinvhierarchy02})
we have neglected $\deltasol$
and $m_1^2$ with respect to \deltaatm{} as well as 
the terms proportional 
to \uuno{} which is limited by the CHOOZ data.
For the best fit value of 
$\deltaatm \cong 3.0\times 10^{-3}~{\rm eV^2}$,
$m_1 < 0.01$ eV and the best fit value of
$\deltasol \cong (5.0 - 5.5)\times 10^{-5}~{\rm eV^2}$,
the corrections due to the $\deltasol$
and $m_1^2$ do not exceed 1\%.
The same conclusion is valid for the
corrections due to the terms
$\sim \uuno $ as long as 
$\uuno \ltap 0.01$. For, e.g,  
$\deltasol \cong 2\times 10^{-4}~{\rm eV^2}$,
$\uuno \cong 0.04$ and $m_1 \cong 0.02$ eV,
the corrections due to the terms neglected
in eq.\ (\ref{eqmassinvhierarchy02}) can reach
$\sim$(5-6)\%. 
These terms, obviously,
should be taken into account if 
they turn out to have the indicated (or larger) 
values and \meff is measured with
a comparable to the corrections, 
or somewhat larger, experimental error. 
In the latter case it will be necessary to use
the exact formula for \meff$\!\!$, 
eq.~(\ref{eqmassinvhierarchy01}),
in order not to introduce avoidable sources
of uncertainties. We are interested here only
i) in analyzing the impact 
for the searches of CP-violation associated
with the Majorana CP-violating phases
of the uncertainty 
in \meff$\!\!$, caused by the uncertainty in the evaluation 
of the relevant \betabeta-decay
nuclear matrix elements,
and ii) in deriving limits on the
nuclear matrix element uncertainty,
which, if satisfied, could permit 
one to draw conclusions concerning 
the CP-violation of interest.  
Therefore in what follows 
we will use the approximate 
expression for \meff$\!\!$,
eq.~(\ref{eqmassinvhierarchy02}),
but our results can be easily generalized
using the exact formula,
eq.~(\ref{eqmassinvhierarchy01}).

   A positive signal for \betabeta-decay in the future
experiments with 
$\sqrt{\deltaatmmin} (\cos 2 \theta_{\odot})_{\mbox{}_{\rm MIN}}
\leq \meff \!\! \leq \sqrt{\deltaatmmax}$,
where \deltaatmmin$\!\!$, \deltaatmmax and 
$(\cos 2 \theta_{\odot})_{\mbox{}_{\rm MIN}}$ 
are determined from the
experimentally measured values 
taking a given C.L. interval,
combined with an upper bound on $m_1$,
$m_1 < 0.02 \ {\rm eV}$,
would lead to the conclusion
that the neutrino mass spectrum is of the 
IH type. A ``just-CP-violating'' region~\cite{BPP1} --- 
a value of \meff in this region would signal unambiguously
CP-violation in the lepton sector
due to Majorana CP-violating phases, would be present if
%%%%%%%%%%%%
\begin{eqnarray}
(\meff_{\rm \!\!exp})_{\mbox{}_{\rm MAX}} <  \sqrt{\deltaatmmin}
\label{suffcondinv01a}
 \\
(\meff_{\rm \!\!exp})_{\mbox{}_{\rm MIN}} >  \sqrt{\deltaatmmax}
( \cos 2 \theta_{\odot} )_{\mbox{}_{\rm MAX}},
\label{suffcondinv01}
\end{eqnarray}
%%%%%%%%%%%%%%%%%%%%%%
%
where $(\meff_{\rm \!\!exp})_{\mbox{}_{\rm MAX ( MIN)}}$
is the largest (smallest) experimentally 
allowed value of \meff$\!\!$,
taking into account both the experimental error
on the measured \betabeta-decay half life-time   
and the uncertainty due to
the evaluation of the nuclear matrix elements.
Condition (\ref{suffcondinv01})
depends crucially on the value of 
$( \cos 2 \theta_{\odot} )_{\mbox{}_{\rm MAX}}$
and it is less stringent for smaller values of 
 $( \cos 2 \theta_{\odot} )_{\mbox{}_{\rm MAX}}$
~\cite{PPW}. 

   We can parametrize the uncertainty in \meff
due to the poor knowledge of the relevant 
nuclear matrix elements --- we will use the term
``theoretical uncertainty'' for the latter,
through a parameter $\zeta$, $\zeta \geq 1$, defined as:
%%%%%%%%%%%%
\begin{equation}
\meff \!\! = \zeta \Big( (\meff_{\rm \!\!exp})_{\mbox{}_{\rm MIN}} \pm
\Delta \Big)
\label{meffpar01}
\end{equation}
%%%%%%%%%%%%%%%%%%%%%%
where $(\meff_{\rm \!\!exp})_{\mbox{}_{\rm MIN}}$ 
is the value of \meff 
obtained from the measured
\betabeta-decay half life-time 
of a given nucleus
using {\it the largest
nuclear matrix element}
and $\Delta$ is  the experimental error.
The necessary condition 
permitting to establish, in principle,
that the CP-symmetry is violated 
due to the Majorana CP-violating phases is:
%%%%%%%%%%%%
\begin{equation}
1 \leq \zeta <  \frac{\sqrt{\deltaatmmin}}
{\sqrt{\deltaatmmax}(\cos 2 \theta_{\odot})_{\mbox{}_{\rm MAX}} + 2 \Delta}~.
\label{invh:cond1a}
\end{equation}
Obviously, the smaller $(\cos 2 \theta_{\odot})_{\mbox{}_{\rm MAX}}$
and $\Delta$ the larger the ``theoretical uncertainty'' 
which might allow one to make conclusions 
concerning the CP-violation of interest.

  From here on, for the sake of simplicity, 
we assume that the ``theoretical uncertainty''
dominates over the experimental one
and we neglect $\Delta$. 
If the computation of the nuclear matrix elements
becomes sufficiently accurate and/or if $\Delta$
is relatively large,
it would be necessary to take into account
also the uncertainty due to the experimental error
in the analysis which follows. 

   For the  best fit value
of \deltaatm{} \cite{Gonza3nu}, 
$(\deltaatm)_{\mathrm{BF}} = 3.1 \times 10^{-3} \ {\rm eV}$,
taking $(\cos 2 \theta_{\odot})_{\mathrm{BF}}= 
0.5, 0.4, 0.3$
and allowing for 10\% (20\%) 
uncertainty in the values of both parameters
\footnote{In order for $\Delta$ to be negligible
in eq.\ (\ref{invh:cond1a}) for 
$\deltaatm = 3.1 \times 10^{-3} \ {\rm eV}$,
$\cos 2 \theta_{\odot} = 0.4$ and an experimental error
of 20\% on both these parameters, one must have 
$\Delta \ll 0.01$ eV. Such a precision in the measurement
of \meff cannot be achieved in the planned 
next generation \betabeta-decay experiments, 
except possibly in the 10 ton version of GENIUS.
Sufficiently small values of $\Delta$ can be achieved
in practically all \betabeta-decay experiments 
of the next generation if $\meff \!\! \gtap 0.20$ eV.}, 
condition (\ref{invh:cond1a}) (with negligible $\Delta$)
implies $\zeta < 1.65, 2.05, 2.73~(1.33,~1.67,~2.22)$, 
respectively.

  Condition (\ref{invh:cond1a})
is obtained for the most favorable case in which
the minimal allowed value of \meff$\!\!$,
$(\meff_{\rm \!\!exp})_{\mbox{}_{\rm MIN}}$,
is very close to the upper bound of the 
allowed range of values for \meff
in the CP-conserving case and opposite CP-parities 
for the two relevant neutrinos,
e.g., for $(\meff_{\rm \!\!exp})_{\mbox{}_{\rm MIN}} =
\sqrt{\deltaatmmax}
( \cos 2 \theta_{\odot})_{\mbox{}_{\rm MAX}}$.
In the general case this condition 
might not be satisfied.
Let us parametrize 
the experimental value of
\meff$\!\!$, $(\meff_{\rm \!\!exp})_{\mbox{}_{\rm MIN}}$,
obtained using {\it the largest 
nuclear matrix element}, as follows:
%%%%%%%%%%%%%%
\begin{equation}
(\meff_{\rm \!\!exp})_{\mbox{}_{\rm MIN}} = 
y \sqrt{\deltaatmmin} (\cos 2 \theta_{\odot})_{\mbox{}_{\rm MIN}},~~~~~~~y \geq 1.
\end{equation}
%%%%%%%%%%%%%%%
Using eq.~(\ref{meffpar01}) we get:
%%%%%%%%%%%%%%
\begin{equation}
\meff \!\! 
= \zeta y \sqrt{\deltaatmmin} (\cos 2 \theta_{\odot})_{\mbox{}_{\rm MIN}}.
\end{equation}
%%%%%%%%%%%%%%%
The requirement that \meff takes values 
in the region allowed in the case 
of neutrino mass spectrum with
inverted hierarchy,
translates into an interval of
allowed values of $y$:
%%%%%%%%%%%%%%
\begin{equation}
1 \leq y \leq  \frac{\sqrt{\deltaatmmax}}{\sqrt{\deltaatmmin}}
\frac{1}{(\cos 2 \theta_{\odot})_{\mbox{}_{\rm MIN}}}.
\end{equation}
%%%%%%%%%%%%%%%
The necessary conditions for establishing 
CP-violation 
due to the Majorana CP-violating phases,
eqs.~(\ref{suffcondinv01a}) and (\ref{suffcondinv01}), 
lead to the following constraints on the 
parameters $y$ and $\zeta$:
%%%%%%%%%%%%%%%%%%
\begin{eqnarray}
\frac{\sqrt{\deltaatmmax}}{\sqrt{\deltaatmmin}}
\frac{(\cos 2 \theta_{\odot})_{\mbox{}_{\rm MAX}}}
{(\cos 2 \theta_{\odot})_{\mbox{}_{\rm MIN}}} < y <
\frac{1}{ (\cos 2 \theta_{\odot})_{\mbox{}_{\rm MIN}}} \label{suffcondinv02}
\\
1 \leq \zeta <  \frac{1}{ y  (\cos 2 \theta_{\odot})_{\mbox{}_{\rm MIN}} }.
\label{suffcondinv03}
\end{eqnarray}
%%%%%%%%%%%%%%%%%%%
The necessary condition for CP- violation
(\ref{invh:cond1a}) (for negligible $\Delta$) can be obtained
from eq.~(\ref{suffcondinv03}) by taking 
$y=\sqrt{\deltaatmmax}(\cos 2 \theta_{\odot})_{\mbox{}_{\rm MAX}}/
(\sqrt{\deltaatmmin}(\cos 2 \theta_{\odot})_{\mbox{}_{\rm MIN}})$.

   In order to exclude the case of CP-conservation
and equal CP-parities of the two relevant neutrinos
$\nu_2$ and $\nu_3$,
the following relation has to be satisfied:
%%%%%%%%%%%%
\begin{equation}
(\meff \!\!)_{\mbox{}_{\rm MAX}} < 
\sqrt{\deltaatmmin},
\end{equation}
%%%%%%%%%%%%%%%%%%%%%%
or, in terms of the parameters $y$ and $\zeta$,
%%%%%%%%%%%%
\begin{equation}
\zeta y <
\frac{1}{ (\cos 2 \theta_{\odot})_{\mbox{}_{\rm MIN}}} .
\end{equation}
%%%%%%%%%%%%%%%%%%%%%%
For $(\cos 2 \theta_{\odot})_{\mbox{}_{\rm MIN}} = 0.40~(0.30)$ 
 and $y=1.0,~1.5,~2.0$,
$\zeta$ needs to satisfy
$\zeta< 2.2,~1.8,~1.2~(3.3,~2.2,~1.67)$.

   The case of CP-conservation
with $\nu_2$ and $\nu_3$ having 
opposite CP-parities,
$\eta_{21}= -\eta_{31} = \pm 1$,
can be excluded if
%%%%%%%%%%%%
\begin{equation}
(\meff\!\!)_{\mbox{}_{\rm MIN}} >  
\sqrt{\deltaatmmax}
 (\cos 2 \theta_{\odot})_{\mbox{}_{\rm MAX}},
\end{equation}
%%%%%%%%%%%%%%%%%%%%%%
or, equivalently  
%%%%%%%%%%%%
\begin{equation}
y > \frac{\sqrt{\deltaatmmax}(\cos 2 \theta_{\odot})_{\mbox{}_{\rm MAX}}}
{\sqrt{\deltaatmmin}(\cos 2 \theta_{\odot})_{\mbox{}_{\rm MIN}}}.
\end{equation}
%%%%%%%%%%%%%%%%%%%%%%
For an uncertainty of 10\% in the values of
\deltaatm{} and $\cos 2 \theta_{\odot}$,
this inequality reads $y > 1.34$. 

  If the neutrino mass spectrum
is of the inverted hierarchy type,
a sufficiently precise determination
of \deltaatm, $\theta_\odot$
and \uuno{} (or a better upper limit on
\uuno), combined with a measurement of \meff
in the \betabeta-decay experiments,
could allow one to get information
on the difference
of the Majorana CP-violating phases
$(\alpha_{31} - \alpha_{21})$
\cite{BGKP96}.
The value of $\sin^2 (\alpha_{31} - \alpha_{21})/2$
is related to the experimentally
measurable quantities as follows \cite{BGKP96,BGGKP99,BPP1}:
%
%%%%%%%%%%%%%%%%%
\begin{equation}
\sin^2 \frac{\alpha_{31} \! - \! \alpha_{21}}{2}
\simeq \Big( 1 - \frac{\meff\!\!^2}{(\deltaatm + m_1^2)\!
(1 - \uuno)^2} \Big) \frac{1}{\sin^2 2  \theta_\odot} \simeq
\Big( 1 - \frac{\meff\!\!^2}{\deltaatm } \Big) \frac{1}{\sin^2 2 \theta_\odot},
\label{CPviolphIH}
\end{equation}
%%%%%%%%%%%%%%%%%
%
where in writing the second simplified 
expression we have assumed that $\uuno \ltap 0.01$ and
$m_1 \ltap 0.01 \ {\rm eV}$. The constraints on
$\sin^2 (\alpha_{31} - \alpha_{21})/2 \neq 0,1$,
which correspond to CP-violation,
are equivalent to eqs.~(\ref{suffcondinv01a}) and 
(\ref{suffcondinv01}). 
Given the fact that the atmospheric neutrino data
implies $\sqrt{\deltaatm} \gtap 0.04$ eV,
obtaining, e.g., an experimental upper limit on \meff
of the order of 0.03 eV would permit,
in particular, to get \cite{PPSNO2bb} 
a lower bound on the value of
$\sin^2 (\alpha_{31} - \alpha_{21})/2$ 
and possibly exclude the CP-conserving case 
corresponding to $\alpha_{31} - \alpha_{21}= 0$
(i.e., $\eta_{21}= \eta_{31}= \pm 1$). 
Note that one of the 
two CP-violating phases,
$\alpha_{21}$ or $\alpha_{31}$, 
will not be constrained
in the case under discussion.
Thus, even if it is found that 
$\alpha_{31} - \alpha_{21}= 0, \pm \pi$,
at least one of the phases
$\alpha_{21}$ and $\alpha_{31}$
can be a source of CP-violation in 
$\Delta L =2$ processes other than
\betabeta-decay.
Let us note also that 
in the limit of negligible (zero) \uuno{} 
and $m_1$, there is practically only 
one physical  CP-violating phase
in the lepton sector in the 
case under discussion
- the Majorana CP-violating phase 
$\alpha_{32} = \alpha_{31} - \alpha_{21}$. 
%%%%%%%%%%%%%%%%%%%%%%%%%%%%%%%%%%%%%%%%%%%%%%%%%%%%%%%%%%%%%%%%%%%%%%%%%%%%%%%
\vspace{-0.3cm}
\subsubsection{Quasi-Degenerate % Neutrino
 Mass Spectrum ($m_1 > 0.2 \ {\rm eV}$, $m_1 \simeq m_2 \simeq m_3 
 \simeq m_{\bar{\nu}_e}$)}
\vspace{-0.2cm}
%%%%%%%%%%%%%%%%%%%%%%%%%%%%%%%%%%%%%%%%%%%%%%%%%%%%%%%%%%%%%%%%%%%%%%%%%%%%%%%

\hskip 1.0cm For the QD neutrino mass spectrum,
the effective Majorana mass \meff is
given in terms of $m_{\bar{\nu}_e} \cong m_{1,2,3}$,
$\theta_\odot$ and $\utre^2$ which is 
constrained by the CHOOZ data, as follows 
(see, e.g., \cite{BPP1,WR00}): 
%%%%%%%%%%%%%%%%%%%%%%%%%%%%%%%%%%%%%%%%%%%%%%
\begin{eqnarray}
\meff \! =  & \me \Big|  \cos^2 \theta_\odot (1 - |U_{\mathrm{e}3}|^2)
+  \sin^2 \theta_\odot (1 - |U_{\mathrm{e}3}|^2) e^{i\alpha_{21}}
+ |U_{\mathrm{e}3}|^2 e^{i\alpha_{31} } \Big|
\label{eqmassdeg01}  \\
\simeq & \me \Big|  \cos^2 \theta_\odot 
+  \sin^2 \theta_\odot  e^{i\alpha_{21}}
 \Big| = \me \sqrt{ 1 - 
\sin^22\theta_{\odot}~\sin^2\frac{\alpha_{21}}{2}}~. 
 \label{eqmassdeg02} 
\end{eqnarray}
%%%%%%%%%%%%%%%%%%%%%%%%%%%%%%%%%%%%%%%%%%%%%%%
%
In eq.~(\ref{eqmassdeg02}) we have neglected \deltasol{} 
and \deltaatm{} since in the case under consideration
$m_1^2 \gg \deltaatm \gg \deltasol$. 
We have furthermore neglected
$ |U_{\mathrm{e}3}|^2$, 
which leads to an uncertainty
on \meff not exceeding 2\% if $\utre^2 < 0.01$. 
If $\utre^2$ turns out to have a value, e.g, 
close to its current 
% 90\% C.L. 
upper limit, $\utre^2 \cong 0.04$, the correction
due to $\utre^2$ in \meff can be as large
as ${\cal O} (8\% \div 12\%)$ and
$\utre^2$ should be kept in the expression for \meff$\!\!$.

  The effective Majorana mass 
\meff in the case of QD spectrum
is limited from below
since $\cos2\theta_{\odot} >0$ 
and the inequality
$m_1^2 \gg \deltaatm$ implies
$m_1 \cong \me \!\! > 0.2 \ {\rm eV}$. 
% (see, e.g., \cite{BPP1}). 
The lower limit on \meff is reached 
in the case of CP-conservation 
and $\eta_{21}=  \eta_{31}= - 1$. 
Using the best fit, 
the $90 \%$ C.L. and the $99.73 \%$ C.L., allowed values,
of $\cos2\theta_{\odot}$ from \cite{SNO2},
we obtain~\cite{PPSNO2bb}, 
respectively, $\meff \!\! \gtap 0.10  \ {\rm eV}$,
$\meff \!\! \gtap 0.06  \ {\rm eV}$ 
and $\meff \!\! \gtap 0.035~{\rm eV}$.

The indicated values of \meff 
are in the range of sensitivity
of some current (NEMO3, CUORICINO) 
and of most future 
\betabeta-decay experiments 
of the next generation.
A measurement of \meff$\!\!$, 
$\meff \!\! > 0.10 \ {\rm eV}$, 
would allow one to conclude that the 
neutrino mass spectrum
is of the QD type.
If, however, \meff is found to lie in the interval
$0.035 \ {\rm eV} \leq \meff \!\! \leq 0.10 \ {\rm eV}$, 
one would need
additional information to establish 
that the neutrino masses are 
quasi-degenerate. For instance,
the inequality
$m_1^2 \cong \me\!\!^2 \gg \deltaatm$,
which for the current best fit value of
\deltaatm{} corresponds to $m_1,\me > 0.2 \ {\rm eV}$,
should be fulfilled. 

  For the values of \meff in the range 
of sensitivity of the discussed
current and future \betabeta-decay experiments, 
a ``just-CP-violation'' region can exist.
In order to establish whether CP-violation
due to the Majorana CP-violating phases takes place, 
the uncertainty on the measured value of \me{} 
should be sufficiently small.
More specifically, the maximal value
of \meff for the allowed range of values of 
\me{} in the case of CP-conservation and 
$\eta_{21} = - \eta_{31}=-1$ 
(see eq.\ (\ref{eqmassdeg01}))
must be smaller than the minimal 
value of \meff for the same
range of \me{} and  $\eta_{21} = - \eta_{31}=1$.
This leads to the following
constraint on the allowed range of 
(or twice the relative error on) \me:
%
%%%%%%%%%%%%%
\begin{equation}
\frac{( m_{\bar{\nu}_e})_{\! \mbox{}_{\rm MAX}} -
( m_{\bar{\nu}_e})_{\! \mbox{}_{\rm MIN}}}
{( m_{\bar{\nu}_e})_{\! \mbox{}_{\rm MAX}}} 
< 1  - (1 + \utremax\!\!)(\cos 2 \theta_{\odot})_{\! \mbox{}_{\rm MAX}}
- \utremax~,
\label{mecond}
\end{equation}
%%%%%%%%%%%%%%%
%
where we have neglected the terms $\sim {\cal O} ( \utremax\!\! )^2$.
Obviously, condition (\ref{mecond})
is less constraining for smaller values of
$(\cos 2 \theta)_{\! \mbox{}_{\rm MAX}}$. 

  If condition (\ref{mecond})
is satisfied, 
the ``just-CP-violation'' interval of values of
$\meff$ is given by:
%%%%%%%%%%%%%%%%%%%%%%%%%%%%%%%
\begin{eqnarray}
(\meff_{\rm \!\!exp})_{\! \mbox{}_{\rm MIN}} > 
\big( (\cos 2  \theta_\odot )_{\! \mbox{}_{\rm MAX}}
(1 - \utremax  \! \! ) + \utremax \!\!\big)
( m_{\bar{\nu}_e})_{\! \mbox{}_{\rm MAX}}~,
\label{CPvioldega}\\
(\meff_{\rm \!\!exp})_{\! \mbox{}_{\rm MAX}}  
 < \big(1 - 2 \utremax \!\!\big)( m_{\bar{\nu}_e})_{\! \mbox{}_{\rm MIN}}.
\label{CPvioldeg}
\end{eqnarray}
%%%%%%%%%%%%%%%%%%%%%%%%%%%%%%%%
%
The necessary condition
for CP-violation, 
eq.~(\ref{CPvioldega}),
is strongly dependent on 
$(\cos 2  \theta_\odot )_{\! \mbox{}_{\rm MAX}}$:
the smaller the value of 
$(\cos 2  \theta_\odot )_{\! \mbox{}_{\rm MAX}}$,
the larger the ``just-CP-violation'' region \cite{PPW}.

   Assuming that $\utre^2 \ltap 0.01$, we can neglect 
$\utre^2$ in eqs.~(\ref{CPvioldega})
and (\ref{CPvioldeg}). Similarly to the case of
IH neutrino mass spectrum, 
one can parametrize
the uncertainty in the value of \meff
associated with the theoretical uncertainty
in the value of the relevant nuclear matrix element(s)
by introducing two real parameters,
$\zeta$ and $y$: $\zeta$ is determined
by eq.\ (10), while
the definition of $y$ in the case under
study can formally be obtained from 
eq.\ (13) by replacing
$\sqrt{\deltaatm}$ with \me. 
The parametrization of \meff$\!\!$, the necessary 
conditions for CP-violation
due to Majorana CP-violating phases,
the constraints on $\zeta$ and $y$, etc.,
can formally be obtained from 
those derived for the case of
IH neutrino mass spectrum in Subsection 3.1.1 
by substituting $\sqrt{\deltaatm}$ with \me,
and we are not going to give them here.
Let us note only that, in general, 
the uncertainty in the measured
value of \me~(or $\Sigma$) 
is expected to be larger than that 
in \deltaatm{} of
${\cal O}(10 \%)$ we have assumed. 
Correspondingly, the former plays a 
more important role as limiting factor 
for the possibility
of detecting the CP-violation 
under discussion (see Section 3.2).
 
  A rather precise determination of \meff$\!\!$, 
$m_1 \cong m_{\bar{\nu}_e}$,
$\theta_{\odot}$ and $|U_{\mathrm{e}3}|^2$ 
would imply an interdependent
constraint on the two CP-violating phases
$\alpha_{21}$ and $\alpha_{31}$~\cite{BPP1,WR00}
(see Fig.\ 16 in \cite{BPP1}).
For $m_1 \equiv m_{\bar{\nu}_e} > 0.2 $ eV, the 
phase $\alpha_{21}$ could be tightly constrained if 
$\utre^2$ is sufficiently small and 
the term in \meff containing 
it can be neglected, 
as is suggested by the current 
limits on $\utre^2$: 
%
%%%%%%%%%%%
\begin{equation}
\sin^2 \frac{\alpha_{21}}{2} \simeq 
\Big( 1 - \frac{\meff\!\!^2}{m_{\bar{\nu}_e}^2} \Big)
 \frac{1}{\sin^2 2\theta_\odot}. 
\label{CPviol1}
\end{equation}
%%%%%%%%%%%%%%%%%%%%%%%%%%%%%%%%
% 
The term  which depends 
on the CP-violating phase $\alpha_{31}$ 
in the expression for \meff$\!\!$,
is suppressed  by the factor $\utre^2$. Therefore 
the constraint one could possibly obtain on 
$\cos \alpha_{31}$ is trivial, 
unless $ \utre^2 \sim {\cal O} ( \sin^2 \theta_\odot)$.
The constraints $\sin^2(\alpha_{21}/2)\neq 0,1$
implying CP-violation, are satisfied if the necessary
conditions for CP-violation,
eqs.~(\ref{CPvioldega}) and (\ref{CPvioldeg}), hold.

     If $\eta_{21}$ and $\eta_{31}$ 
take the CP-conserving values  
$\eta_{21}= \pm \eta_{31} = - 1$,
there are both an upper and a 
lower bounds on \meff$\!\!$,
$m_{\bar{\nu}_\mathrm{e}} 
((\cos 2  \theta_\odot )_{\! \mbox{}_{\rm MIN}} ( 1 - \utremin\!\!)
+ \utremin\!\!) \leq \meff
\leq m_{\bar{\nu}_\mathrm{e}}
((\cos 2  \theta_\odot )_{\! \mbox{}_{\rm MAX}} ( 1 - \utremax\!\!)
+ \utremax\!\!)$, where we have used eq.\ (\ref{eqmassdeg01}).
Given the range of allowed values 
of $\cos2\theta_{\odot}$,
the observation of the \betabeta-decay in 
the present and/or future 
\betabeta-decay experiments,
combined with a sufficiently 
stringent upper limit on 
$m_{\bar{\nu}_e} \simeq m_{1,2,3}$,
~$m_{\bar{\nu}_e} < 
\meff_{\rm \!\! exp} / ((\cos 2  \theta_\odot )_{\! \mbox{}_{\rm MAX}}
(1 - \utremax  \! \! ) + \utremax \!\!)$,
would permit, e.g., to exclude
the case of CP-conservation with 
$\eta_{21}=\pm \eta_{31}= - 1$ \cite{PPSNO2bb}.
%
%%%%%%%%%%%%%%%%%%%%%%%%%%%%%%%%%%%%%%%%%%%%%%%%%%%%%%%%%%%%%%%%
\vspace{-0.3cm}
\subsection{Example of Simplified Error Analysis}
\vspace{-0.2cm}
%%%%%%%%%%%%%%%%%%%%%%%%%%%%%%%%%%%%%%%%%%%%%%%%%%%
%
\hskip 1.0cm  In this subsection we give an example
of simplified error analysis of prospective data on 
$m_{\bar{\nu}_e}$ or $\Sigma$,
$\tan^2\theta_{\odot}$, \meff$\!\!$, etc.,
performed with the aim of establishing 
at a given C.L. that the phases
$\alpha_{21}$ and/or $\alpha_{31}$, or
$\alpha_{32} = \alpha_{31} - \alpha_{21}$,
take CP-violating values.
For simplicity we use eq.\ (\ref{eqmassdeg02})
(eq.\ (\ref{eqmassinvhierarchy02})), 
assuming that 
$\sin^2\theta \equiv \utre^2~(\uuno)$
is sufficiently small, $\sin^2\theta \ltap 0.01$.
In this case we get 
for the CP-violating phase of interest
either eq.\ (\ref{CPviol1}) (QD spectrum) or
eq.\ (\ref{CPviolphIH}) (IH spectrum).

   In the analysis which follows
we use the following 4 representative 
values of  $\tan^2 \theta_\odot$ from the region of the LMA solution:
$\tan^2 \theta_\odot = 0.25;~0.40;~0.55;~ 0.70$. 
The 1 s.d.\ error of the experimentally
measured value of $\tan^2 \theta_\odot$ is assumed to be 
$\sigma (\tan^2 \theta_\odot)/\tan^2 \theta_\odot = 5\%$.
Considering the case of QD neutrino mass spectrum,
we take the following illustrative values of
$m_0^2 \equiv \me\!\!^2 \cong m_{1,2,3}^2$
(to be measured in the \hbeta experiment
KATRIN) and $\Sigma$ (which
could be determined from cosmological
and astrophysical data):
$m_0^2 = (1.0)^2;~(0.70)^2;~(0.50)^2~{\rm eV^2}$
and $\Sigma = 3.0;~1.5;~0.60$ eV, 
with $m_0 = \Sigma/3$. The error on the value of 
$m_0^2$ used in the analysis is 
$\sigma (m_0^2) = 0.08~{\rm eV^2}$ \cite{KATRIN},
while that on the value of
$\Sigma$ is assumed to be 
$\sigma(\Sigma) = 0.04~{\rm eV}$ \cite{Hu99}.
The latter implies for the QD spectrum that
$\sigma (m_0) = \sigma (\Sigma)/3 \cong 0.013$ eV. 
We represent the error of the 
experimentally measured value of \meff in the 
standard form:
%
%%%%%%%%%%%%%%%%%%%%%%%%%%%%%%%%%%%%%
\begin{equation}
  \frac{\sigma (\meff\!\!)}{\meff} = \sqrt{ (E_1)^2 + (E_2)^2}~,
\label{Emeff}
\end{equation}
%%%%%%%%%%%%%%%%%%%%%%%%%%%%%%
% 
where $E_1$ and $E_2$ are the statistical and systematic
errors. We choose $E_2 = const = 0.05$. 
We take $E_1 = f/\meff\!\!$, where we assume $f$ = 0.028 eV.
This gives a total relative error 
$\sigma (\meff\!\!)/\meff \!\! \cong 15\%$ at $\meff \!\! = 0.20$ eV.
The above choices are motivated by the fact that
the sensitivities of the next generation of 
\betabeta-decay experiments 
(CUORE, GENIUS, EXO, MAJORANA, MOON)
in the measurement of \meff
are estimated to be in the range
of $\sim (1.5 - 5.0)\times 10^{-2}$ eV and if, e.g,
$\meff \!\! \gtap 0.20$ eV, a precision in the determination
of \meff corresponding to an error of $\sim 15\%$
could be reached in these experiments. 
Moreover, for values of \meff which are 
sufficiently bigger than the quoted 
sensitivity limits of the
future experiments, the statistical error
scales as \meff increases like $E_1 \sim const/\meff\!\!$.
 
  The measurement of \meff and $m_0^2$ and the
more accurate determination of $\tan^2 \theta_\odot$ 
would allow one
to determine $\sin^2\alpha$, where $\alpha \equiv 
\alpha_{21}/2$ for the QD case, 
using eq.\ (\ref{CPviol1}). 
Using error multiplication, the error on $\sin^2\alpha$ is:
%
%%%%%%%%%%%%%%%%%%%%%%%%%%%%%
\begin{eqnarray} % \label{eq:sigal1}
\sigma (\sin^2 \alpha) = \frac{\D (1 + \tan^2 \theta_\odot)^2}{\D 4 \; \tan^2 \theta_\odot} \left[
4 \left( \frac{\D \meff\!\!^2}{\D m_0^2}\right)^2 \; 
\left(\frac{\D \sigma(\meff\!\!)}{\meff}\right)^2 + 
\left( \frac{\D \meff\!\!^2}{\D m_0^2}\right)^2 \; 
\left(\frac{\D \sigma (m_0^2)}{\D m_0^2}\right)^2 \right. \nonumber \\[0.4cm]
\left. + \left(\frac{\D 1 -  \tan^2 \theta_\odot}{\D 1 + \tan^2 \theta_\odot}\right)^2 
\left(1 - \frac{\D \meff\!\!^2}{\D m_0^2} \right)^2 
\left(\frac{\D \sigma (\tan^2 \theta_\odot)}{\D \tan^2 \theta_\odot}\right)^2 \right]^{1/2} \; .
\label{Esin2alpha}
\end{eqnarray}
%%%%%%%%%%%%%%%%%%%%%%%%%%%%%%%%%%
If the sum $\Sigma = 3m_0$ is cosmologically determined,
one has $\sigma(\Sigma)/\Sigma = \sigma(m_0)/m_0$ and 
the error on $\sin^2 \alpha$ can be obtained from 
the above equation by using 
$\sigma (m_0^2)/m_0^2 = 2\sigma (m_0)/m_0$.

  In deriving eq.\ (\ref{Esin2alpha})
we have assumed that there are no correlations 
between the errors on $m_0^2$ (or $m_0$), $\tan^2 \theta_\odot$ and \meff$\!\!$. 
Let us note that the first two terms 
in the expression for $\sigma (\sin^2 \alpha)$
are suppressed for $m_0^2 \gg \meff\!\!^2$, i.e.,
for $\alpha \sim \pi/2$. The third is small for 
$\meff \!\! \simeq m_0$, i.e., for $\alpha \sim 0,\pi$,
when the first two dominate, 
and/or for $\tan^2 \theta_\odot \simeq 1$. 
The common factor 
in eq.\ (\ref{Esin2alpha}) is minimized for $\tan^2 \theta_\odot = 1$.

    With the help of eq.\ (\ref{Esin2alpha}) one can study 
the dependence of the error of $\sin^2 \alpha$ on the 
values of the neutrino mass and mixing 
parameters and their errors. Figures 
1 and 2 show results for a possible 
determination of a neutrino mass from 
the tritium spectrum and cosmology, respectively. 

  As Figs.\ 1 and 2 demonstrate, 
establishing under the assumptions made
that CP-violation takes place, i.e., that
$\sin^2(\alpha_{21}/2) \neq 0,1$,
at  99.73\% C.L.
requires typically $\tan^2 \theta_\odot \gtap 0.40$,
$m_0^2 \gtap 0.70^2~{\rm eV^2}$,
or $\Sigma \gtap 1.0$ eV. Further, the value of 
the CP-violating phase 
$\alpha_{21}$  should 
lie approximately within the intervals
$\sim (\pi/2 - 3\pi/4)$ and
$\sim (5\pi/4 - 3\pi/2)$. 
The larger the values of $\tan^2 \theta_\odot$, and/or $m_0^2$
or $\Sigma$, the larger the interval of values of 
$\alpha_{21}$ for which
CP-violation could be established.

    Next we include the effect of the uncertainty in the
relevant \betabeta-decay nuclear matrix elements 
in the analysis by assuming that the value of \meff
which enters into the expression for     
$\sigma (\sin^2 \alpha)$, eq.\ (\ref{Esin2alpha}),
is obtained from a measurement of the \betabeta-decay 
lifetime of given nucleus by using 
{\it the maximal allowed value of the nuclear matrix element},
i.e., that
$\meff \!\! = \zeta \Big( (\meff_{\rm \!\!exp})_{\mbox{}_{\rm MIN}}\Big)$,
where $\zeta \geq 1$ parametrizes the uncertainty 
under discussion (see subsection 3.1.1).
Figures 1 and 2 correspond then to  
$\zeta = 1$. In Figures 3 -- 6 we exhibit
results for $\zeta = 1.5;~2.0;~3.0$, including in 
each sub-figure also the results for $\zeta = 1$.
As Figs.\ 3 -- 6 show, it would be impossible to make 
a definite conclusion concerning the CP-violation
for values of the parameters considered,
$m_0^2 \equiv  m_{\bar{\nu}_e}^2 = (0.5^2 - 1.0^2)~{\rm eV^2}$, 
$\Sigma = (0.60 - 3.0)$ eV, $\tan^2 \theta_\odot = (0.25 - 0.70)$,
if $\zeta \gtap 3$. Actually, for
$\tan^2 \theta_\odot \sim 0.25$, $m_0^2 \sim (0.8^2 - 1.0^2)~{\rm eV^2}$
or $\Sigma \sim (1.5 - 3.0)$ eV, one might even exclude the
possibility of $\zeta \gtap 3$ at 99.73\% C.L.
Establishing at $3\sigma$ that the phase
$\alpha_{21}$ takes a CP-violating value 
is possible provided $\zeta < 2$,
$\tan^2 \theta_\odot \gtap 0.55$,
$m_0^2 \gtap 0.70^2~{\rm eV^2}$, or
$\Sigma \gtap 1.5$ eV, and if 
$\alpha_{21} \sim (\pi/2 - 3\pi/4)$ or
$\alpha_{21} \sim (5\pi/4 - 3\pi/2)$. 
The precise sub-intervals of values of
$\alpha_{21}$ for which CP-violation
could be established depend of 
the precise values of $\tan^2 \theta_\odot$,
$m_0^2$, or $\Sigma$. 
For $\zeta = 2$ this might be done
as well, but
for a rather limited range of 
values of $\alpha_{21}$
from the indicated intervals
and if $\tan^2 \theta_\odot \gtap 0.65$,
$m_0^2 \gtap 0.80^2~{\rm eV^2}$, or
$\Sigma \gtap 2.0$ eV.
Obtaining an evidence for CP-violation
at $2\sigma$ level for a given
$\zeta \ltap 2$ is possible for 
wider ranges 
of values of $\alpha_{21}$, 
$\tan^2 \theta_\odot$, $m_0^2$, or $\Sigma$,   
than those permitting 
a $3\sigma$ ``proof''.

  Similar results are valid for neutrino mass 
spectrum with inverted hierarchy. In this case the role of
$m_0^2$ is played by $\deltaatm$ 
(see eq.\ (\ref{eqmassinvhierarchy02})),
which is expected to be measured with a relative error of 
$\sim 10\%$. Reaching a definite conclusion 
concerning the CP-violation,
as the preceding discussion indicates,
requires, in particular, \meff to be measured 
with an error not exceeding $\sim (15 - 20)\%$.

Once \meff$\!\!$, $\theta_\odot$ and $m_{\bar{\nu}_e}$
are measured in present and future experiments, 
constraints at a given C.L. on the allowed values of $\zeta$ 
and $\sin^2 \alpha_{21}/2$ can be obtained 
performing a joined $\chi^2$ analysis of the
data on \meff$\!\!$, $\theta_\odot$ and $m_{\bar{\nu}_e}$.

%%%%%%%%%%%%%%%%%%%%%%%%%%%%%%%%%
\vspace{-0.3cm}
\section{Conclusions}
\vspace{-0.2cm}
%%%%%%%%%%%%%%%%%%%%%%%%%%%%%%%%%

\hskip 1.0truecm Assuming 3-$\nu$ mixing and massive 
Majorana neutrinos, \betabeta-decay generated 
only by the (V-A) charged current weak interaction 
via the exchange of the three Majorana neutrinos,
LMA MSW solution of the $\nu_{\odot}$-problem and
neutrino oscillation explanation of the  
atmospheric neutrino data, 
we have discussed in the present article 
the possibility of 
detecting CP-violation in the lepton sector, 
associated with Majorana neutrinos, from a 
measurement of the effective Majorana mass 
in \betabeta-decay, \meff$\!\!$.
The problem of detection of CP-violation 
in the lepton sector  
is one of the most formidable and challenging 
problems in the studies of neutrino mixing.
As was noticed in \cite{PPW}, 
the  measurement of \meff alone
could exclude the possibility of 
the two Majorana CP-violating phases 
$\alpha_{21}$ and $\alpha_{31}$,
present in the lepton mixing matrix 
in the case of interest, 
being equal to zero. However,
such a measurement cannot rule out
without additional input
that the two phases take different 
CP-conserving values.
The additional input needed for
establishing CP-violation
could be, e.g., the measurement of 
neutrino mass $m_{\bar{\nu}_e}$ in 
\hbeta experiment
KATRIN \cite{KATRIN}, or the 
cosmological determination of
the sum of the three neutrino masses \cite{Hu99},
$\Sigma = m_1 + m_2 + m_3$,
or a derivation of a sufficiently 
stringent upper limit on $\Sigma$ or
on the lightest neutrino mass $m_1$.
At present no viable alternative to 
the measurement of 
\meff for getting information 
on the Majorana CP-violating phases 
$\alpha_{21}$ and $\alpha_{31}$
exists, or can be foreseen 
to exist in the next $\sim 8$  years.
Thus, the present work represents
an attempt to determine the 
conditions under which 
CP-violation might be detected  
from a measurement of 
\meff and of $m_{\bar{\nu}_e}$ (or $\Sigma$), or
 of \meff and by obtaining a sufficiently 
stringent upper limit on $\Sigma$ or $m_1$.

  We have discussed in detail the
prospective data on the neutrino mass, 
mixing and oscillation parameters on which 
\meff depends: $m_{\bar{\nu}_e}$, $\Sigma$,  \deltaatm,
$\tan^2\theta_{\odot}$,  \deltasol, \uetre or \uuno. 
Considering neutrino mass spectrum with inverted hierarchy
and of quasi-degenerate type, we performed 
a general analysis,
the aim of which was to determine what is the
maximal uncertainty in the value of \meff due to the
imprecise knowledge of the corresponding 
\betabeta-decay nuclear matrix
elements, which might still allow one to
find CP-violation associated 
with Majorana neutrinos.
This was followed % in Section 3.2 
by an example of simplified error analysis 
of the possibility to establish that 
the physical Majorana 
phases take CP-non-conserving values.
The analysis is based on
prospective input data on 
\meff$\!\!$, $m_{\bar{\nu}_e}$, $\Sigma$,
$\tan^2\theta_{\odot}$, etc.
The effect of the nuclear matrix element
uncertainty was included in the 
analysis. The results thus obtained are 
illustrated in Figs.\ 1 -- 6.

  The possibility of finding the CP-violation 
of interest requires quite accurate measurements
of \meff and, say, of $m_{\bar{\nu}_e}$ or $\Sigma$, 
and holds only for a limited range of 
values of the relevant parameters.
More specifically, 
proving that CP-violation associated with
Majorana neutrinos takes place
requires, in particular, a relative 
experimental error on the measured value of 
\meff not bigger than (15 -- 20)\%,
a ``theoretical uncertainty'' in the value of
\meff due to an imprecise knowledge of the 
corresponding nuclear matrix elements
smaller than a factor of 2, a value of 
$\tan^2\theta_{\odot} \gtap 0.55$,
and values of the relevant Majorana
CP-violating phases ($\alpha_{21}$, 
$\alpha_{32}$) typically 
within the ranges of $\sim (\pi/2 - 3\pi/4)$ and
$\sim (5\pi/4 - 3\pi/2)$. 

\vspace{0.2cm}
 {\bf Acknowledgments.} S.T.P. would like to thank 
H. Ejiri and G. Gratta for useful discussions.
The work of W.R. has been supported by the
``Bundesministerium f\"ur Bildung, Wissenschaft, Forschung und
Technologie'', Bonn under contract No.\ 05HT1PEA9.
The work of S.T.P. was supported in part by 
the EC network HPRN-CT-2000-00152.

\begin{figure}
\epsfig{file=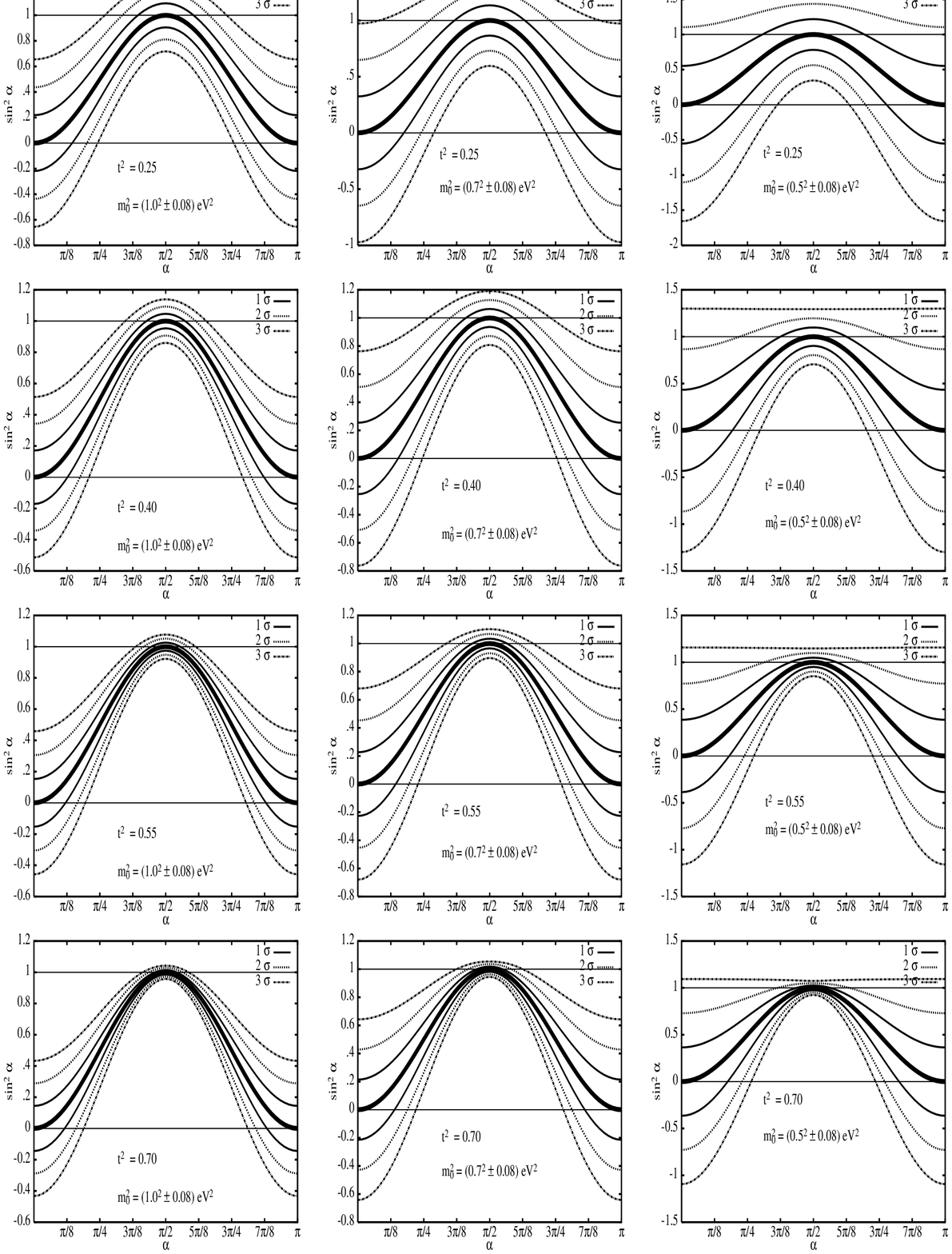,width=16cm,height=21cm}
\caption{The error on $\sin^2 \alpha$ as a function of
 $\alpha \equiv \alpha_{21}/2$ 
 in the case of QD neutrino mass spectrum
 (see eq.\ (\ref{CPviol1}))
 for different values of $m_0^2 \equiv \me\!\!^2$ and 
 $t^2 \equiv\tan^2  \theta_\odot$.
 The neutrino mass parameter  
 $m_0^2 \equiv \me\!\!^2 \equiv m_{1,2,3}^2$ 
 is assumed to be determined 
 in the \hbeta experiment KATRIN \cite{KATRIN}
 with an error of  0.08 eV$^2$.
 The error on $\tan^2 \theta_\odot$ is
 5$\%$, while that on \meff is given by 
 eq.\ (\ref{Emeff}). 
 The 1 $\sigma$ range is within the solid lines, the 
 2$\sigma$ (3$\sigma$) error band is within the 
 dashed (dash-dotted) lines.}
\end{figure}

\begin{figure}
\epsfig{file=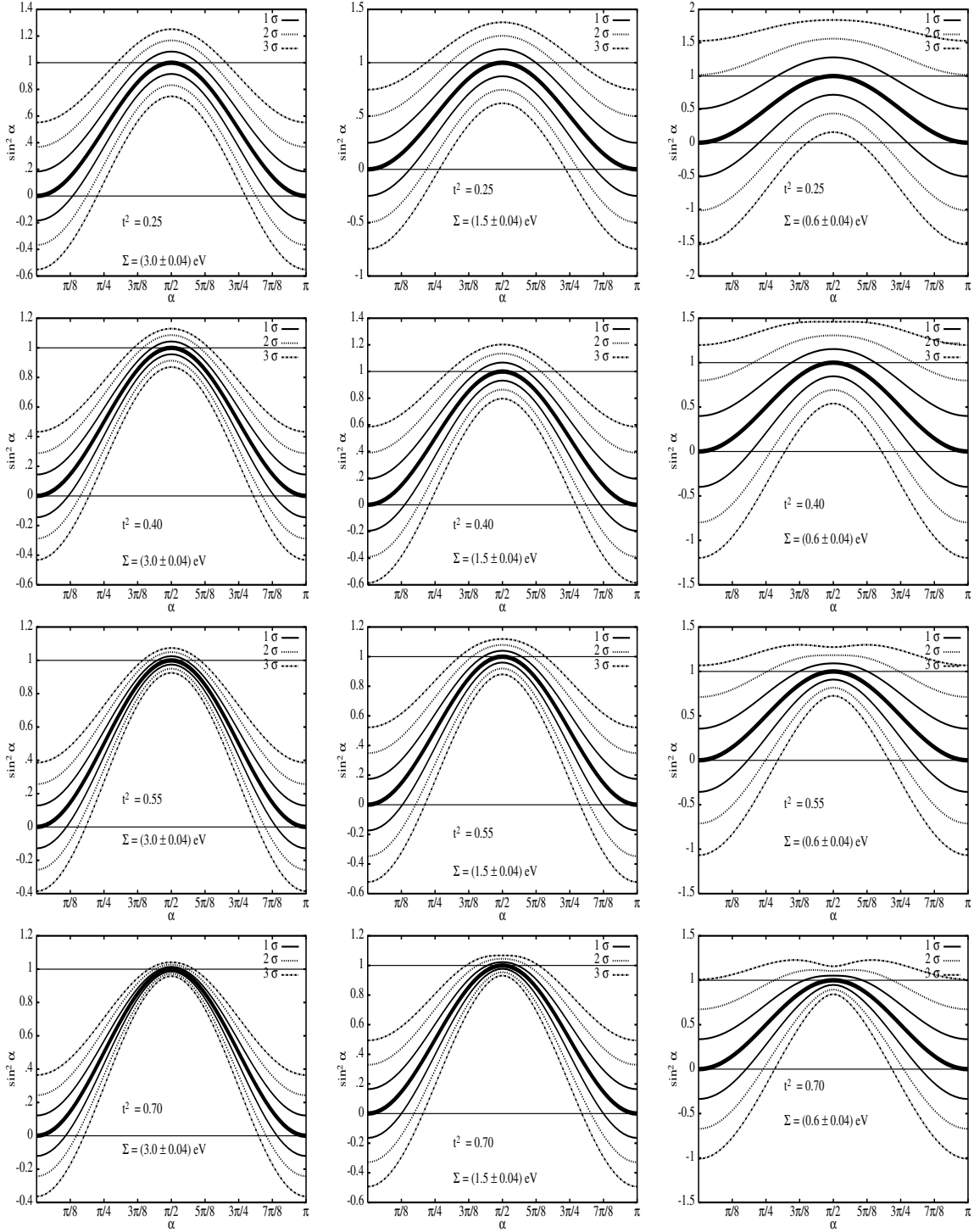,width=16cm,height=21cm}
\caption{The error on $\sin^2 \alpha$ as a function of
 $\alpha \equiv \alpha_{21}/2$ in the case 
 of QD neutrino mass spectrum
 (see eq.\ (\ref{CPviol1}))
 for different values of 
 $\Sigma = 3m_0$ ($m_0 \equiv \me$) and  $t^2 \equiv\tan^2  \theta_\odot$.
 The sum of the neutrino masses
 $\Sigma$ is assumed to be determined 
 from astrophysical and cosmological
 data with an error of 0.04 eV \cite{Hu99}. 
 The error on $\tan^2 \theta_\odot$
 is 5\%, while the error 
 on \meff is given by eq.\ (\ref{Emeff}).
 The 1 $\sigma$ range is within the solid lines, 
 the 2$\sigma$ (3$\sigma$) error band is within the
 dashed (dash-dotted) lines.}
\end{figure}

\begin{figure}
\epsfig{file=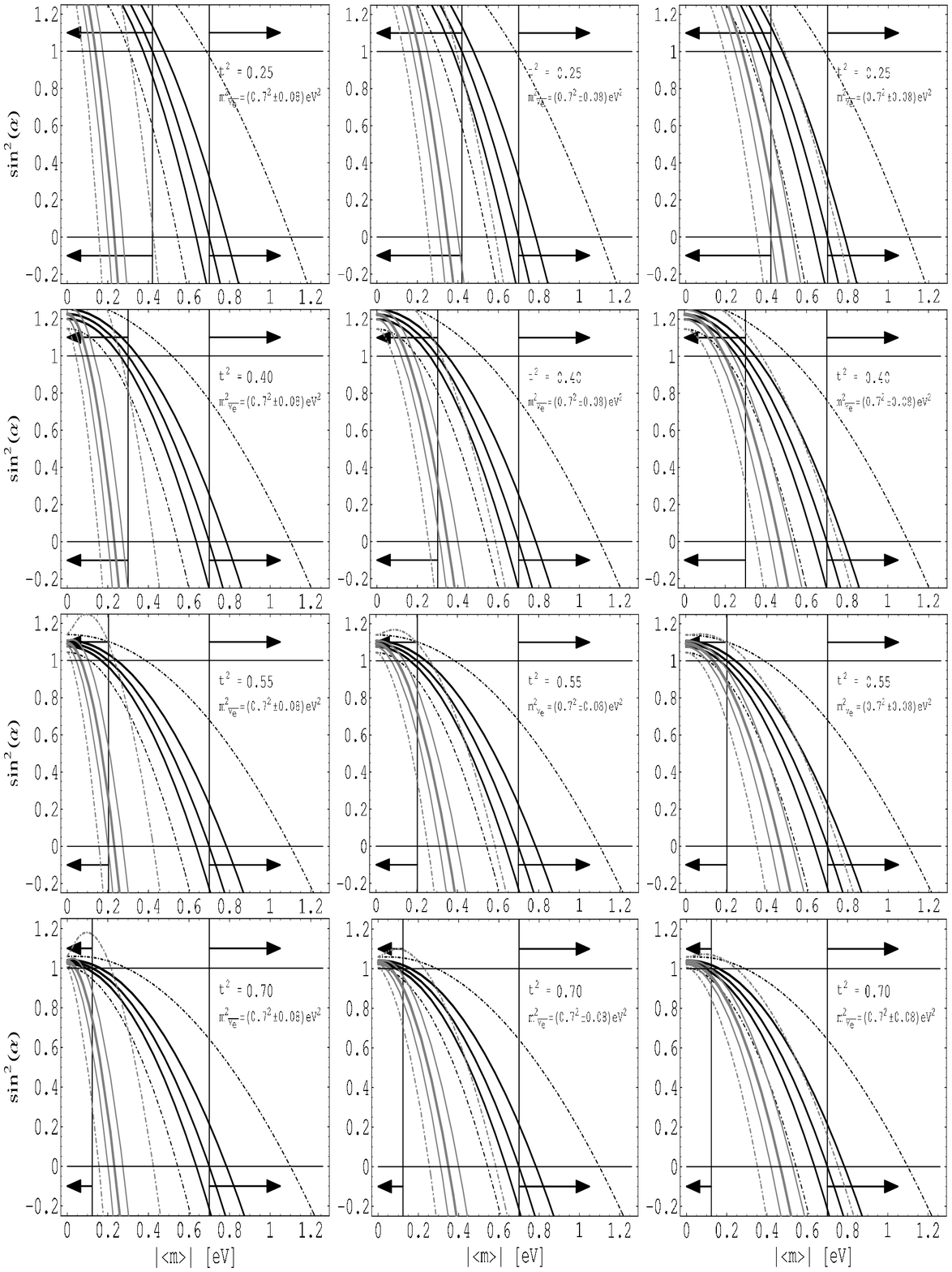,width=16cm,height=21cm}
\caption{ The same as in Fig.\ 1 for  
 $m_0^2 = \me\!\!^2 = (0.70^2 \pm 0.08)~{\rm eV^2}$ 
 and three values of nuclear matrix element
 uncertainty factor $\zeta$: 3.0 (left panels),
 2.0 (middle panels) and 1.5 (right panels).
 The 1 $\sigma$ range shown for $\zeta = 1.5;~2.0;~3.0$
 ($\zeta = 1.0$) 
 is within the light-gray (black) solid lines, the 
 3$\sigma$ error band is within the 
 light-gray (black) dash-dotted lines. 
 The arrows denote the  values of \meff
 which lie outside the allowed region 
 in the case of the QD neutrino mass spectrum.
}
\end{figure}

\begin{figure}
\epsfig{file=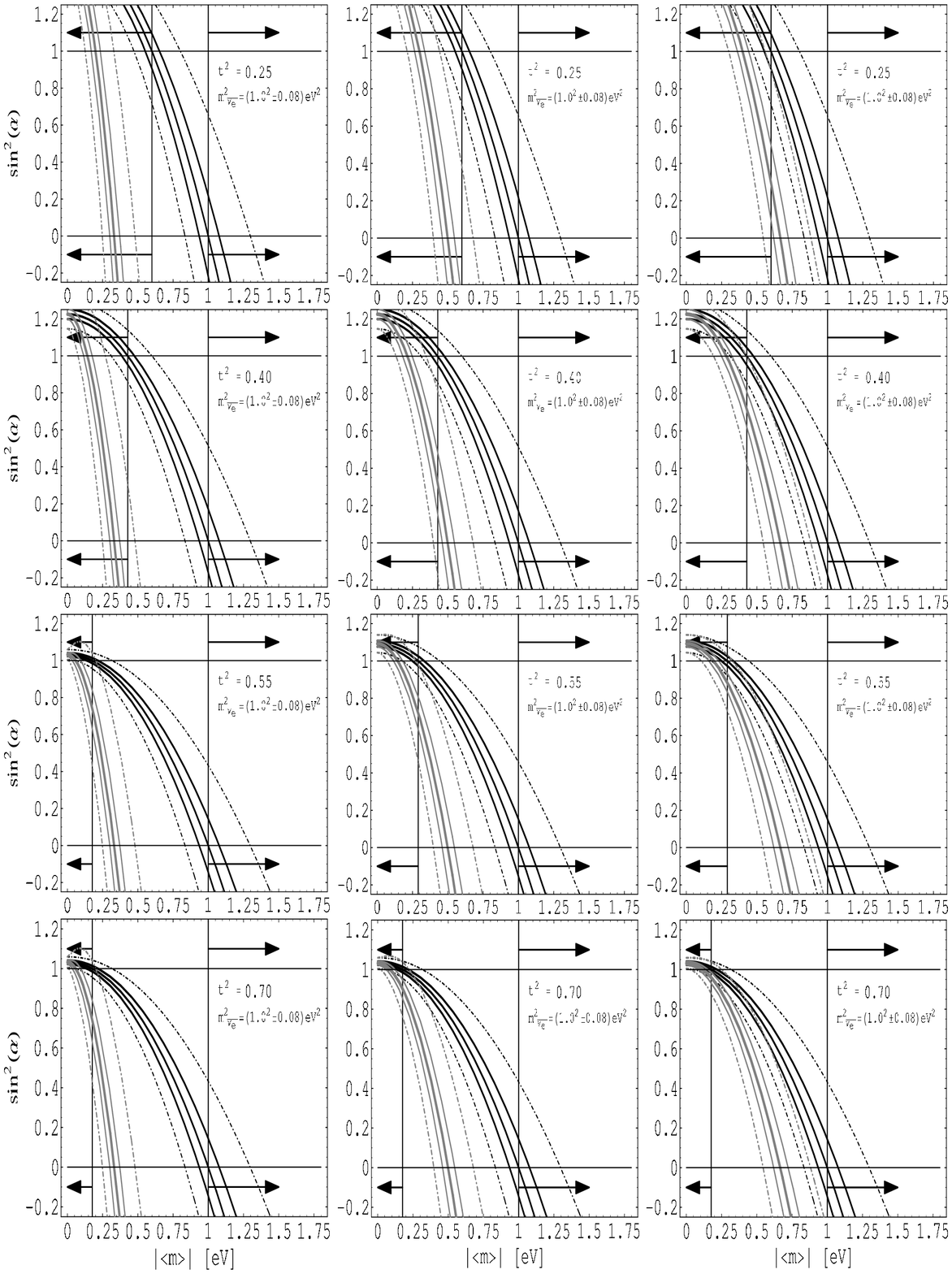,width=16cm,height=21cm}
\caption{The same as in Fig.\ 1 for 
$m_0^2 =\me\!\!^2 = (1.0^2 \pm 0.08)~{\rm eV^2}$ 
 and three values of nuclear matrix element
 uncertainty factor $\zeta$: 3.0 (left panels),
 2.0 (middle panels) and 1.5 (right panels).
 The 1 $\sigma$ range shown for $\zeta = 1.5;~2.0;~3.0$
 ($\zeta = 1.0$) 
 is within the light-gray (black) solid lines, the 
 3$\sigma$ error band is within the 
 light-gray (black) dash-dotted lines.  
 The arrows denote the  values of \meff
 which lie outside the allowed region 
 in the case of the QD neutrino mass spectrum.
}
\end{figure}

\begin{figure}
\epsfig{file=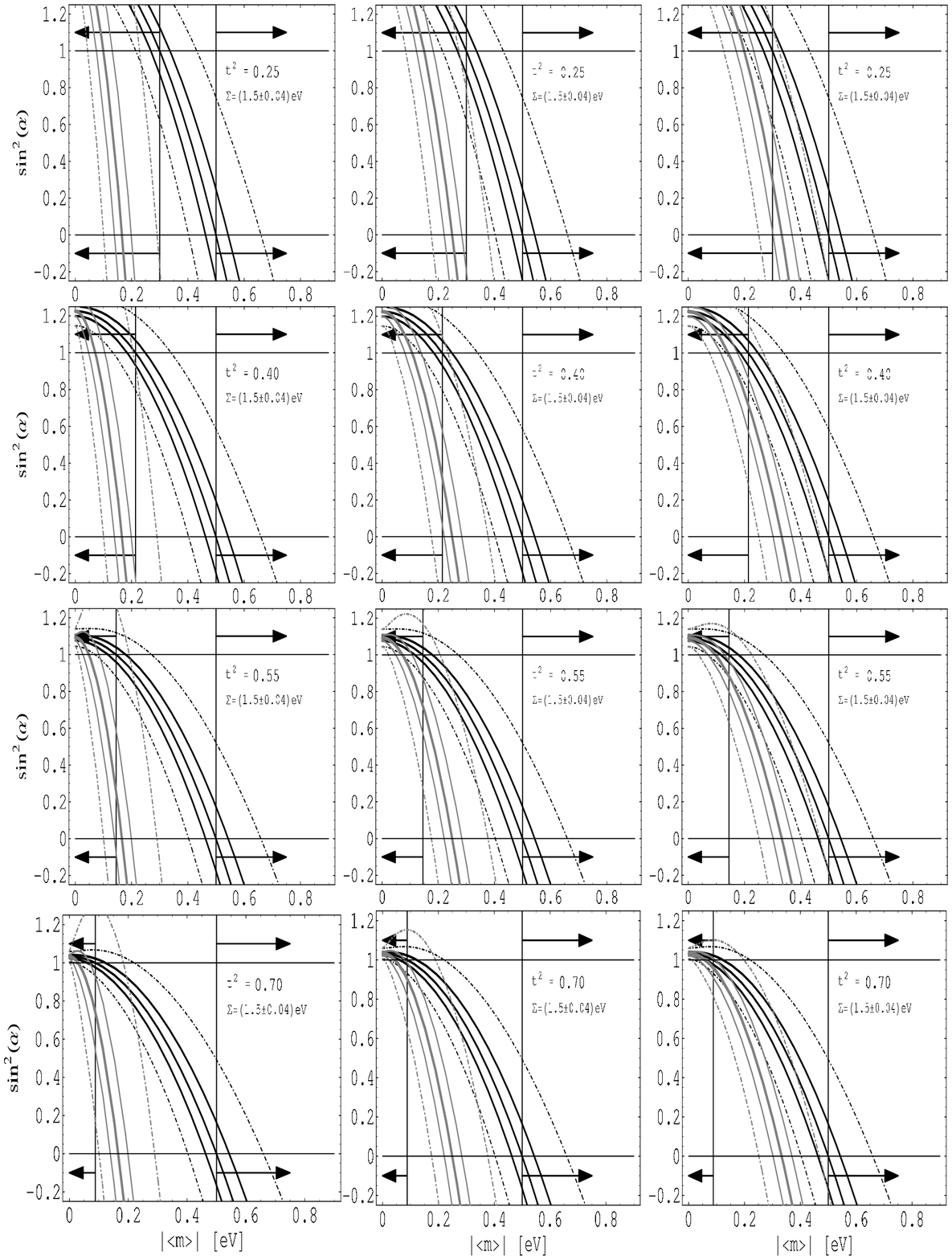,width=16cm,height=21cm}
\caption{The same as in Fig.\ 2 for  
 $\Sigma = (1.5 \pm 0.04)~{\rm eV}$ 
 and three values of nuclear matrix element
 uncertainty factor $\zeta$: 3.0 (left panels),
 2.0 (middle panels) and 1.5 (right panels).
 The 1 $\sigma$ range shown for $\zeta = 1.5;~2.0;~3.0$
 ($\zeta = 1.0$) 
 is within the light-gray (black) solid lines, the 
 3$\sigma$ error band is within the 
 light-gray (black) dash-dotted lines.
 The arrows denote the  values of \meff
 which lie outside the allowed region 
 in the case of the QD neutrino mass spectrum.
}
\end{figure}

\begin{figure}
\epsfig{file=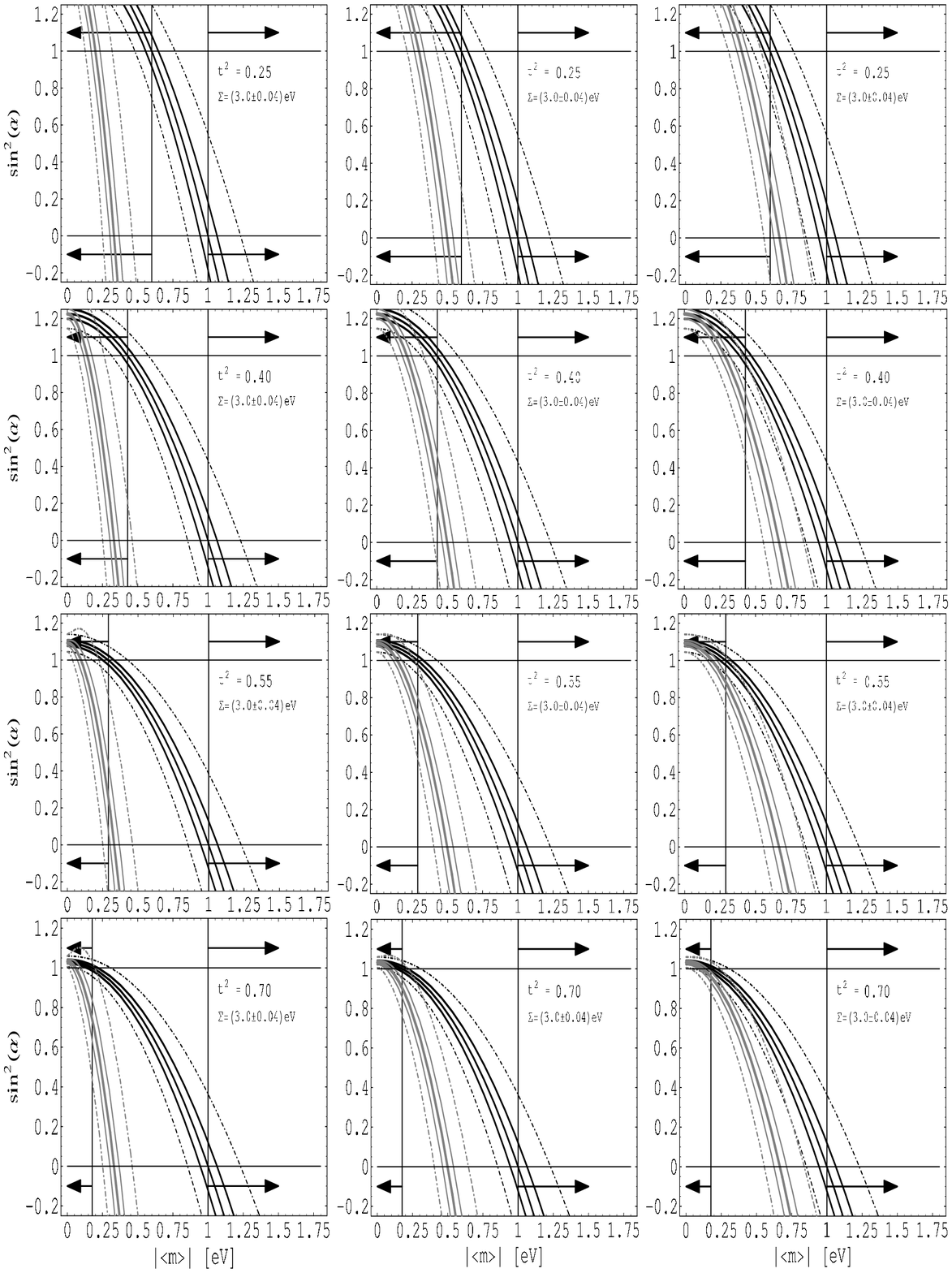,width=16cm,height=21cm}
\caption{The same as in Fig.\ 2 for  
  $\Sigma = (3.0 \pm 0.04)~{\rm eV}$ 
 and three values of nuclear matrix element
 uncertainty factor $\zeta$: 3.0 (left panels),
 2.0 (middle panels) and 1.5 (right panels).
 The 1 $\sigma$ range shown for $\zeta = 1.5;~2.0;~3.0$
 ($\zeta = 1.0$) 
 is within the light-gray (black) solid lines, the 
 3$\sigma$ error band is within the 
 light-gray (black) dash-dotted lines.
 The arrows denote the  values of \meff
 which lie outside the allowed region 
 in the case of the QD neutrino mass spectrum.
}
\end{figure}

\end{document}